\documentclass[12pt,preprint]{aastex}

\newcommand{\simgt}{\hbox{\rlap{\raise 0.425ex\hbox{$>$}}\lower 0.65ex\hbox{$\sim$}}}
\newcommand{\simlt}{\hbox{\rlap{\raise 0.425ex\hbox{$<$}}\lower 0.65ex\hbox{$\sim$}}}

\newcommand{\tth}{{\twothirds}}
\newcommand{\p}{\,^\prime}
\newcommand{\pp}{\,^{\prime\prime}}

\newcommand{\powerlaw}{$\rho/\sigma^3$}
\newcommand{\facto}{scaling parameter}
\newcommand{\factos}{scaling parameters}  

\begin{document}
\title{\Large {Semi-analytical dark matter halos and the Jeans equation}}
\pagestyle{plain} 
\textheight8.95in

\author{Crystal G. Austin, Liliya L.R. Williams, Eric I. Barnes}
\affil{Astronomy Department, University of Minnesota, Minneapolis, MN 55455} 
\email{caustin,llrw,barnes@astro.umn.edu} 

\author{{Arif Babul}\altaffilmark{1}}
\affil{Department of Physics and Astronomy, University of Victoria, BC, Canada} 
\email{babul@uvic.ca} 
\altaffiltext{1}{Leverhulme Visiting Professor, Universities of Oxford and Durham}

\author{{Julianne J. Dalcanton}\altaffilmark{2}} 
\affil{Astronomy Department, University of Washington, Box 351580, Seattle, WA 98195} 
\email{jd@astro.washington.edu} 
\altaffiltext{2}{Alfred P. Sloan Research Fellow}

\begin{abstract}
Although N-body studies of dark matter halos show that the density profiles,
$\rho(r)$, are not simple power-laws, the quantity $\rho/\sigma^3$, where
$\sigma(r)$ is the velocity dispersion, is in fact a featureless power-law 
over $\sim 3$ decades in radius. In the first part of the paper we 
demonstrate, using the semi-analytic Extended Secondary Infall Model (ESIM),  
that the nearly scale-free nature of $\rho/\sigma^3$ is a robust feature of 
virialized halos in equilibrium. By examining the processes in common between 
numerical N-body and semi-analytic approaches, we argue that the scale-free 
nature of $\rho/\sigma^3$ cannot be the result of hierarchical merging, 
rather it must be an outcome of violent relaxation. The empirical results of 
the first part of the paper motivate the analytical work of the second part 
of the paper, where we use $\rho/\sigma^3\propto r^{-\alpha}$ as an additional
constraint in the isotropic Jeans equation of hydrostatic equilibrium. Our 
analysis shows that the constrained Jeans equation has different types of 
solutions, and in particular, it admits a unique ``periodic'' solution with 
$\alpha=1.9444$. We derive the analytic expression for this density profile,
which asymptotes to inner and outer profiles of $\rho\sim r^{-0.78}$, and 
$\rho\sim r^{-3.44}$, respectively.
\end{abstract}

\section{Introduction}

There is a broad consensus that gravity-driven cosmological evolution 
of collisionless cold dark matter (CDM), starting from some realistic
matter power spectrum, results in virialized objects whose 
spherically-averaged density profile is well represented by the NFW prescription
~\citep{nfw97}, or its close variants \citep{m98,nav04}. The main
characteristics of the density profile are the relatively shallow
density slope in the center, $d\log\rho/d\log r\sim-1$, and much steeper 
slope around the virial radius, $\approx -3$. These general properties 
are well established, but the details are not. The value of the central
slope, and the nature of the transition between inner and outer slopes
is still a matter of debate.  Currently, some advocate the original NFW
formulation with constant asymptotic density slopes and a scale-radius
\citep{diemand}. Others argue for a more continuous change of 
the slope as a function of radius \citep{nav04,merritt}. Regardless 
of which density profile expression proves to best describe N-body 
halos, the underlying physics that drives the halos to have this shape 
is not yet fully understood. 

Several attempts have been made in the literature to explain the density 
profile shape. For example, \cite{sw98} suggested that the NFW profile 
results from a series of repeated mergers; as the undigested cores of 
merging satellites accumulate at the center, the central $r^{-1}$ cusp 
develops. However, \cite{huss} performed simulations of isolated halos with 
varying degrees of thermal motion of dark matter particles. The final halos 
closely resembled the universal NFW shape, suggesting that hierarchical 
merging is not critical to the outcome. \cite{m99} simulated halo 
formation using a matter power spectrum with suppressed small scale power. 
The final density profiles were still well described by the NFW fit.
Some analytical studies suggest that secondary infall and accretion can 
determine the density profile shape \citep{avila}. \cite{barnes05a} argue 
that the density profile, and in particular its radially changing slope, 
is the result of a mild aspect of the radial orbit instability. 

A different approach was taken by \citet{tn01} (hereafter TN01), and 
\cite{h04}. Instead  of considering the radial run of the space density 
of N-body halos, they measured the quantity $\rho(r)/\sigma^3(r)$, which 
has the dimensionality of phase-space density. In spherically-symmetric 
equilibrium halos, \powerlaw\/ is proportional to the coarse-grained 
phase-space density, a quantity distinct from the (fine-grained) 
phase-space density whose conservation is ensured by the collisionless
Boltzmann equation \citep{dehnen05}. As our analysis does not hinge on 
the strict interpretation of the quantity \powerlaw, we will continue 
to call it phase-space density, for brevity. 
 
TN01 found that \powerlaw\/ of their three halos (circular velocities
160-200 km\,s$^{-1}$) is very well approximated by a single 
power-law, from $10^{-2.5}$ virial radii to just beyond the virial radius. 
This finding is unexpected because the density profile, $\rho(r)$ 
undergoes a considerable slope change in the same radial interval. 
The scale-free nature of the phase-space density implies that the double 
logarithmic slope of the velocity dispersion changes in such a way so as 
to offset the change in the density profile slope. The power-law 
index $\alpha$, where \powerlaw$\propto r^{-\alpha}$, was found to be 
very similar for all the three halos studied by TN01; $\alpha=1.875$.

Does the power-law nature of \powerlaw\/ characterize all equilibrium N-body
halos, or are the halos studied by TN01 somehow special? One can get a rough
answer to this question without analyzing many individual simulation halos. 
Given the form of a ``universal'' empirical fit to N-body halos and assuming that 
the halos are in equilibrium, one can calculate the phase-space density as a 
function of radius using the Jeans equation.  \cite{barnes05b} use two types 
of empirical fits to N-body profiles \citep{nfw97,nav04} to demonstrate that 
both produce nearly scale-free phase-space density profiles for $\sim 3$ 
decades in radius, with a slope close to 1.9. They conclude that most 
N-body halos should have scale-free phase-space density profiles. This 
conclusion is further supported by a recent study of equilibrium halos 
\citep{kazan04}. 
The halo that most resembles results
of N-body simulations has scale-free \powerlaw\/ and index $\alpha\approx1.95$.

The scale-free nature of \powerlaw\/ represents a novel way of looking at 
the properties of halos, and as such, should be explored further.  If this 
property is ``universal'', it amounts to a hitherto unrecognized constraint 
on the shape of the density profiles. In this paper we address questions 
related to the \powerlaw\/ profiles of collisionless halos. 
Is the scale-free nature of \powerlaw\/ (i) a result of initial conditions, 
(ii) an outcome of evolution and virialization, or (iii) a consequence of the 
final equilibrium state of halos?  In a related work, \cite{barnes05b} 
address the third question. They demonstrate that the condition of 
hydrostatic equilibrium\footnote{We use the phrase `hydrostatic equilibrium' 
because collisionless dark matter can be thought of as fluid, with 
$\rho\sigma^2$ acting as pressure \citep{shap04}. }  
alone, as quantified by the Jeans 
equation, does not require \powerlaw\/ to be a power-law in radius.

The first two questions can be addressed by studying halo formation
techniques different from N-body simulations. Semi-analytical 
methods based on the secondary infall paradigm are an example of 
such techniques. 
Halos forming in N-body simulations are subject to a complex set of 
dynamical processes. Particles can exchange energy and momentum 
through violent relaxation and two-body encounters 
(even though the latter are suppressed in cosmological simulations). 
Additionally, phase-mixing can decrease phase-space density, while 
preserving particles' energies. These processes are brought about 
by halo collapse, merging, dynamical friction, etc. The analytical 
methods do not incorporate such a wide range of processes. The
method we use here is based on \cite{RG87}, and was extended by 
\cite{wbd04}; we call it ESIM, the Extended Secondary Infall Model.
As we describe later in more detail, ESIM halos are spherically symmetric 
with one spatial and two velocity (radial and tangential) phase-space 
dimensions. They virialize through violent relaxation, which changes
the shells' energies, but conserves the angular and radial momenta 
of each shell throughout the collapse. There are no mergers.
  
Given such different formation scenarios, it is not surprising that the 
final properties---density distributions, etc.---of N-body and ESIM halos
tend to be different. (See \cite{barnes05a} for a discussion
of why the two types of density profiles differ.)
However, if specific final properties are present in N-body and ESIM halos,
that strongly suggests that the property is the result of shared physics. 
This is our premise in addressing question (ii). Specifically, we find 
that the phase-space density of our halos is well represented by a power-law
(Section~\ref{ESIMsec}), a result similar to the one reached through the 
study of numerically generated halos.
Question (i) is also addressed in Section~\ref{ESIMsec}, where we vary
input conditions of ESIM halos. Again, the power-law nature of \powerlaw\/ does 
not seem to be affected by these indicating that the answer to question
(i) is negative. 

In Sections~\ref{jeans} we address the behavior of index $\alpha$. One of the 
main properties of the virialized halos is that they are in equilibrium, so 
the Jeans equation is important in determining the structure of halos. 
Therefore it makes sense to ask if the Jeans equation has a preferred value 
for $\alpha$, or, if all values are equally likely from the point of view 
of the Jeans equation. Our analysis of the second derivative of the Jeans 
equation and its various solutions leads to some interesting insights
(also see Appendix~\ref{convex}). Appendix~\ref{SIM} derives the value of 
$\alpha$ for halos formed in a highly idealized secondary infall model (SIM), 
and compares it to $\alpha$'s obtained in ESIM halos of Section~\ref{ESIMsec}.

\section{Extended Secondary Infall Model (ESIM) Halos}\label{ESIMsec}
\subsection{Summary of the Method}

The collisionless dark matter halos in the present study are formed
using an analytical method originally introduced by \cite{RG87}.
The method, including minor extensions, is fully described in
\cite{wbd04}. In summary, halos are spherically symmetric and isolated.  
We use a flat, matter dominated geometry with $h=0.5$. The initial power 
spectra is a standard CDM Harrison-Zeldovich power spectrum normalized 
such that $\Omega^{0.6}\sigma_8=0.5$.  Evolution of the power spectrum 
is dictated by the transfer function given by \cite{BBKS} in their 
Appendix G.  The density profile of a proto-halo is taken to be the
profile of a peak in a density field described by this power spectrum, 
$P(k)$ as is illustrated in \cite{wbd04}, Figure 9. 

In addition to the 
primary peak, proto-halos have superimposed secondary perturbations, 
whose properties are derived from the same $P(k)$. In a real halo these 
secondary perturbations generate random motions of dark matter particles, 
both radial and tangential. In ESIM halos, the RMS amplitude of random 
velocities (as a function of radius) are calculated from $P(k)$, as 
described in \cite{RG87}, and in the Appendix of \cite{wbd04}. For each 
shell the magnitude of the velocity perturbation is chosen from a 
Maxwell-Boltzmann distribution with the above RMS; the direction is
chosen randomly. It has been recently pointed out that the 
Maxwell-Boltzmann distribution does not describe the particle velocities 
of equilibrium halos \citep{kazan04,h05}. However, the Maxwell-Boltzmann
distribution is sufficient for our purposes, because we use it to describe
the initial pre-collapse, not the final equilibrium halos. Even if the
Maxwell-Boltzmann assumption is not strictly correct for the proto halos,
it suffices for our purposes, because we use ESIM halos as a simplified
tool only; we do not aim to reproduce real halos.

When a particular shell reaches its maximum turn-around radius the 
perturbation in radial and tangential velocity is added to the existing 
radial infall velocity. In a real halo these additional random velocities 
result in dark matter particles having elliptical, precessing orbits. 
In ESIM halos, the random velocities transform the initially thin shells 
into thick, overlapping shells, each with its own apo- and peri-center. A key 
assumption in ESIM calculations is that the collapse proceeds adiabatically, 
with each shell conserving its angular and radial momentum. Note that
each shell's secondary velocity perturbation is oriented randomly, so that 
the halo as a whole has zero net angular momentum, and no net rotation. 
As the collapse proceeds, the mass distribution in the halo changes, and 
the potential and shells' energies are recalculated after each step in the 
collapse. This is violent relaxation. To sum up, ESIM halos are formed via 
smooth accretion of somewhat lumpy material, with no major mergers.

Our method of halo formation has considerable flexibility with direct
control over the parameter space of initial conditions. For example,
we can change the shape of the primary density peak by smoothing or
altering the matter power spectrum. Both the shape of the primary peak 
and the amplitude of the secondary perturbations are based on $P(k)$,
but we can alter the shape of the primary peak and the amplitude of
secondary perturbations independently, which gives us another degree of 
freedom in exploring the resulting final halo profiles. For example, we can
change the secondary perturbations---the overall amplitude and radial 
dependence---while keeping the primary peak unaltered. Even though these 
changes need not correspond to any astrophysically plausible scenarios,
the evolution of ESIM halos proceeds in a physically consistent way.

\subsection{ESIM Halo Calculation Results}\label{ESIMhalos}

We generate three sets of halos, with `galaxy', `group' and `cluster' scale
masses. 
To get different ranges of halo mass we filter the mass power spectrum, 
prior to normalization, on comoving scales $r_h$ of 0.05, 0.35 and 0.7 Mpc, 
respectively. 
Larger $r_h$ mean that the collapse will start later, and more massive 
halos will form as a result. Within each mass set we study the effect of the 
secondary velocity perturbations on the final halo profiles. To that 
end, we form halos where the magnitudes of the secondary velocity 
perturbations are scaled by a parameter, hereafter called the perturbation
amplitude \facto, or \facto\/ for short. 
In this Section the \facto\/ is altered uniformly for a given halo, but in 
Section~\ref{resiliency} we radially vary it within the halos.  A \facto\ 
of zero corresponds to no secondary perturbations, in which case the 
velocities are purely radial. A \facto\ of 1 corresponds to the original 
amplitude of secondary perturbations derived from the same power spectrum 
that determined the shape of the primary density peak, while setting the 
\facto\/ to 2 doubles the amplitude of all secondary perturbations.

The perturbation amplitude \facto\/ is varied from 0.4 to 2.5 in increments 
of 0.05.  Because each shell is assigned a 
velocity perturbation from a distribution extending to 
$\approx 2.5 V_{rms}$, there 
is a chance that the shell's kinetic energy will exceed its potential 
energy. This is especially likely if the \facto\/ is large, and the shell
happens to be far from center.  In this case, we assume that the shell 
is instantaneously lost from the halo and is no longer considered in the 
potential calculations or the kinematic analysis. 
Each shell's potential energy is calculated based on the mass distribution
due to a finite number of shells, with the outermost shells extending well 
beyond the virialized portion of the halo at the 
present epoch. The total number of shells, and hence the zero-point of 
the potential, is fixed such that halos computed using \factos\/ less 
than 1.5 do not lose any mass during the collapse, while 
halos with a \facto\/ of 2 lose at most 25\% of their final mass. 

Each ESIM halo is unique since the secondary velocities for each shell
are picked randomly from a distribution.  To reduce the
dispersion in the final halo properties the halos presented here are
averages of ten individual halos. To verify the reliability of the results
two sets of averaged halos were computed for each \facto.
In all cases our analysis applies to the portion of each halo between 
$10^{-2.5}$ and 1 virial radius, $r_{200}$, the radius at which the
average enclosed halo density is 200 times the critical, 
$\rho_{o}=2.77\times 10^{11} h^2 M_\odot\;Mpc^{-3}$. As mentioned 
earlier, we use a standard cold dark matter model. While this model
is not the currently favored one, the particular cosmological model 
used has a minimal effect on the final halo properties of interest.

In the following sections we describe the various properties of the final 
virialized ESIM halos. In Sections~\ref{ESIMdensity}-\ref{mass} we discuss
in detail the properties of galaxy mass halos. Since the qualitative
properties and the general trends are similar for galaxy, group, and 
cluster mass halos, we do not discuss the latter two specifically, but
present a summary of their properties in Section~\ref{rgvalues}.

\subsubsection{Density profiles}\label{ESIMdensity}

\begin{figure}[t]
\plotone{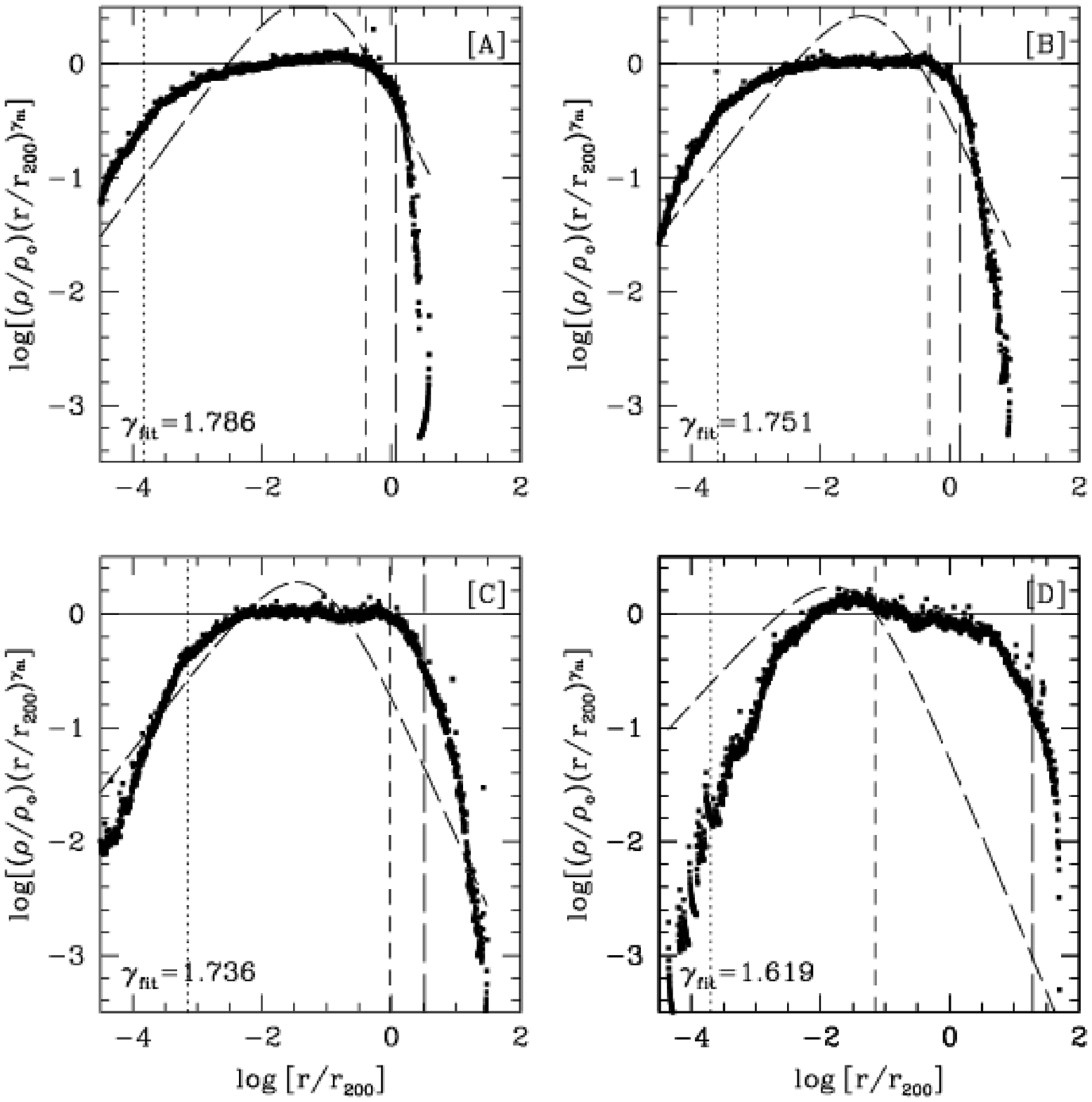}
\caption{
      Density vs. radius in units of virial radius, $r_{200}$, for galaxy-mass 
      halos generated with $r_h=0.05$Mpc, and perturbation amplitude \factos\/  
      of 0.4 [A], 0.6 [B], 1 [C], and 2 [D]. Density is expressed in units of 
      the critical density, and multiplied by $(r/r_{200})^{\gamma_{fit}}$,
      to accentuate deviations from a power-law. $-\gamma_{fit}$, the best
      power-law fit to the density profile between $10^{-2.5}$ and 1 virial 
      radii, is noted in each panel. The vertical normalization in this plot
      is arbitrary. The dotted, short dash and long dash lines 
      represent radii where the density slope is 1, 2  and 3; $r_{\gamma=1,2,3}$.
      The NFW profile is over-plotted as a long dash line in all 4 panels.  
      The NFW concentration parameters were computed using the masses of the 
      corresponding halos.  
	}
\label{density}
\end{figure}

\begin{figure}[t]
\plotone{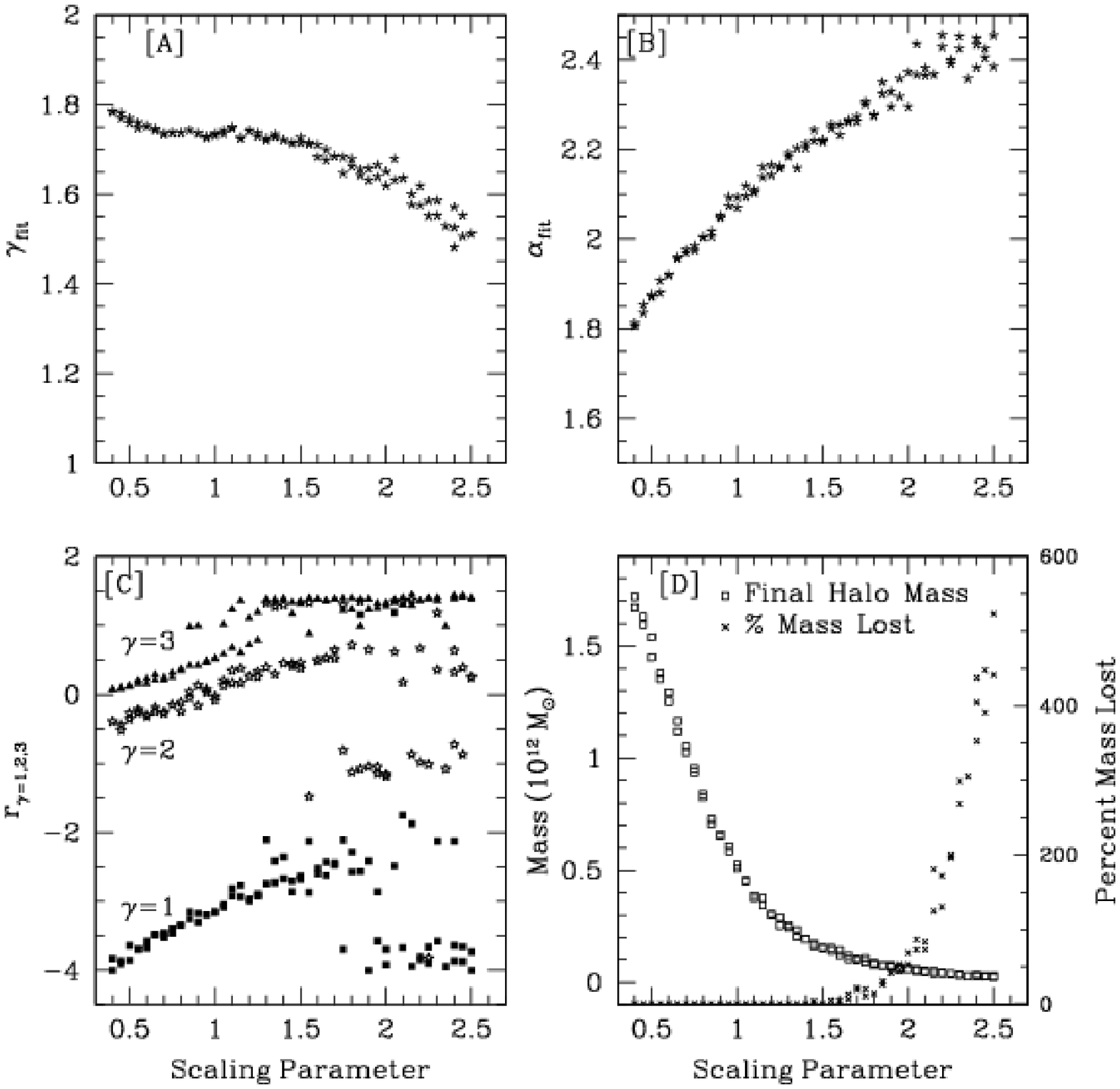}
\caption{
	The final properties of galaxy-mass halos with $r_h=0.05$Mpc as
        as a function of the perturbation amplitude \facto. 
	[A]: Slope of the log-log density profile, $\gamma_{fit}$;
	[B]: Slope of the log-log \powerlaw\/ profile, $\alpha_{fit}$;
	[C]: $r_{\gamma=1}$, $r_{\gamma=2}$, and $r_{\gamma=3}$: radii, in 
	terms of virial, where the double logarithmic density slopes attain 
        values of 1, 2, and 3, respectively;
        [D]: Virial mass of the halo (squares, and left vertical axis) and 
        percentage mass lost during formation, with respect to the final
        halo mass (crosses, and right vertical axis).
        Note that as the amplitude of secondary perturbations is increased, the 
        resulting halos have shallower density slopes ([A] and [C]) due to increased 
        angular momentum, steeper phase-space density slopes [B], and more mass loss [D]
        due to the increased probability of matter reaching escape velocity.
} 
\label{vsfacto}
\end{figure}

Figure~\ref{density} shows four halo density profiles, expressed in units of 
critical density $\rho_{o}$, and multiplied by $(r/r_{200})^{\gamma_{fit}}$,
where $-\gamma_{fit}$ is slope of the best power-law fit to the density
profile between $10^{-2.5}r/r_{200}$ and $r/r_{200}$. This is done in order 
to accentuate features in the density profiles. All halos were generated using 
the galaxy-mass $P(k)$ filtering scale $r_h=0.05$Mpc. The standard case, a 
halo with a \facto\/ of 1, is shown in panel C. Panels A, B and D show halos 
with \facto\/ of 0.4, 0.6 and 2, respectively. 

As in \citet{wbd04}, the ESIM halos differ from those in N-body simulations. 
For convenience, NFW density profiles, 
${\rho(r)}\propto{(r/r_s)(1+r/r_s)^2}$ \citep{nfw97}, with concentration 
parameter ($c=r_{200}/r_s$) appropriate for that halo mass, and arbitrary
normalization are over-plotted as long dash lines.  
ESIM density profiles are better approximated by a single 
power-law between $r=10^{-2.5}r_{200}$ and $r_{200}$ than are NFW profiles. 

The effects of changing the amplitude of secondary perturbations on the mass 
distribution in the halos can be seen in the four panels of Figure~\ref{density},
and is summarized in panel A of Figure~\ref{vsfacto}, which plots
$\gamma_{fit}$, the fitted density profile slopes, vs. \facto. As expected,
the halos with larger secondary perturbations, and hence more angular
momentum have shallower density profiles. (Note that the steepest central 
cusp slope of ESIM halos is $\gamma=2$, as explained in Section 5.1 of
\citet{wbd04}.) To further characterize the density profiles we indicate in 
Figure~\ref{density} where the slope, $\gamma_{fit}=-(d\ln\rho)/(d\ln r)$, attains 
the values of $\gamma$=1, 2, and 3, with dotted, short-dash and long dash lines.
We denote these radii by $r_{\gamma=1}$, $r_{\gamma=2}$, and $r_{\gamma=3}$.  The
dependence of $r_{\gamma=1,2,3}$ with respect to the scaling parameter is shown 
in panel D of Figure~\ref{vsfacto}. The values of $r_{\gamma=1,2,3}$ 
increase with the \facto, consistent with density profiles becoming shallower as 
the angular momentum content of the halos increase.

\subsubsection{Velocity dispersion and velocity anisotropy}

\begin{figure}[t]
\plotone{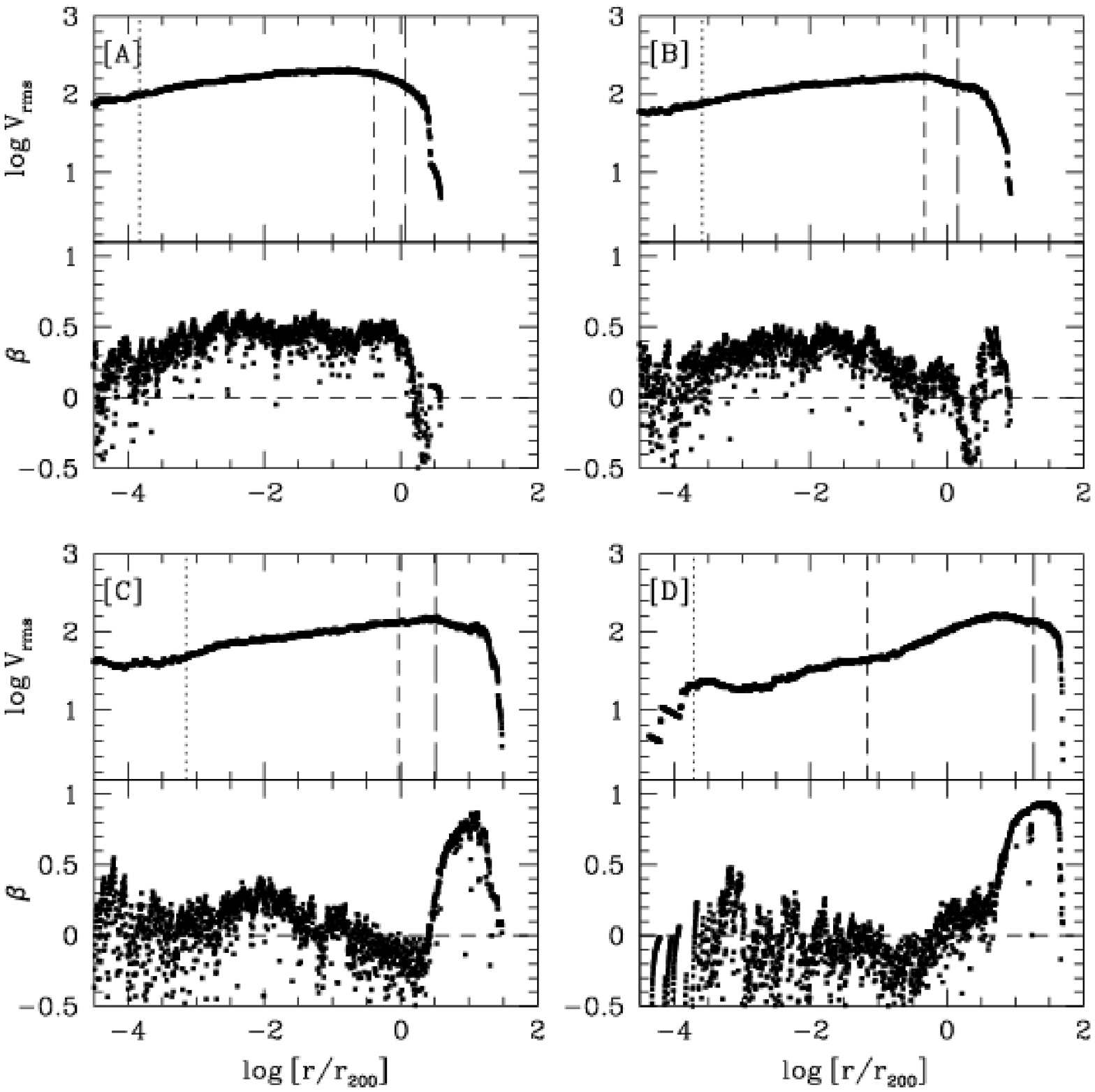}
\caption{
	The top half of each of the four panels shows the RMS velocity profile for
        halos with perturbation amplitude \factos\/ of 0.4 [A], 0.6 [B], 1 [C], 
        and 2 [D]. The lower half of each panel shows the velocity dispersion 
	anisotropy, $\beta$.  Again, the values of $r_{\gamma=1,2,3}$ are shown in 
        dotted, short dash, and long dash lines. Halos [A] and [B] have somewhat
        radial velocity anisotropies, while halos [C] and [D] have nearly isotropic
	velocity dispersions up to the virial radius.
	} 
\label{velocity}
\end{figure}

Since many shells overlap at any given radius, the total velocity dispersion
profile is calculated as the mass-weighted and time-weighted RMS value of the 
radial and tangential velocity dispersions of all the shells that exist at 
that radius. Figure \ref{velocity} shows the velocity dispersion profiles for
the halos presented in Figure~\ref{density}. The values of $r_{\gamma=1}$, 
$r_{\gamma=2}$, and $r_{\gamma=3}$ are marked as dotted, short-dash and 
long-dash lines. Figure \ref{velocity} also shows the velocity dispersion 
anisotropy, $\beta$, defined as usual, through the radial and tangential 
velocity dispersions \citep{BT},
\begin{equation}
\beta\equiv1-\frac{\sigma_\theta^2}{\sigma_r^2}.
\label{betaell}
\end{equation}
Except for halos generated using small \facto\/ values, $\simlt\;1$, the 
anisotropy parameter stays close to 0 between $\sim 10^{-3}$ and $\sim 1$ 
virial radii. For the standard halo, the average velocity anisotropy 
parameter is $\beta \sim 0.1$.  Beyond the virial 
radius, the velocity ellipsoid parameter increases to $0.5-1$ as the orbits 
become more radial.  In general, a halo's average $\beta$ 
increases with decreasing  \facto: smaller secondary perturbations imply
less angular momentum, and hence more radial orbits. In the limit of
perturbation amplitude \facto\/ of 0, purely radial orbits will result.

\begin{figure}[t]
\plotone{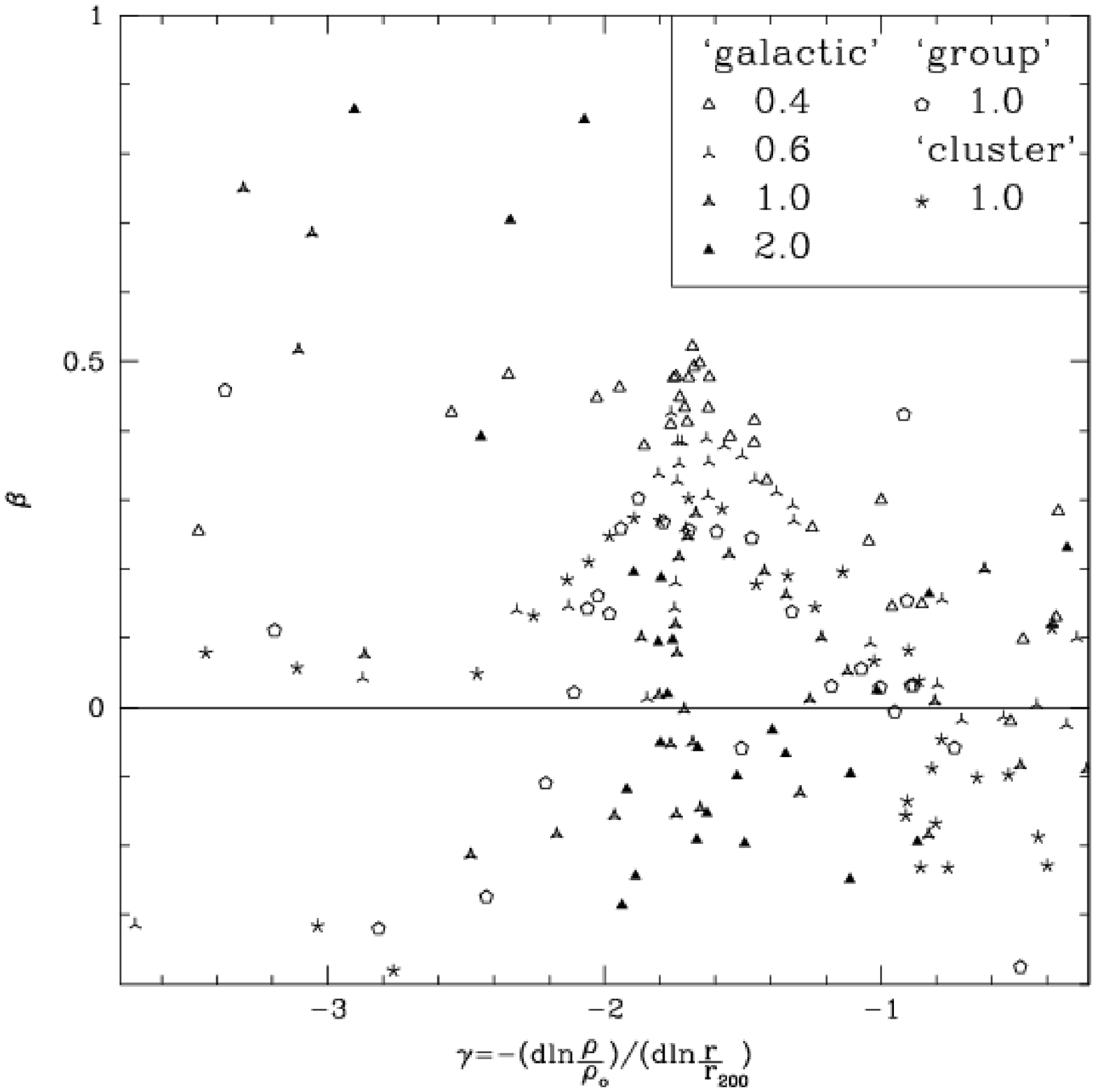}
\caption{The anisotropy parameter $\beta$ vs. the density slope $\gamma_{fit}$ 
      relation for ESIM halos with 
      \factos\/ of 0.4, 0.6, 1, and 2.  Each set of points represents a 
      halo; each point represents a different radial section of that halo.  
      We also display the results of the `group' and `cluster' sized halos 
      discussed in Section~\ref{rgvalues}.
	} 
\label{betavsgamma}
\end{figure}

In Figure~\ref{betavsgamma}, we show the dependence of the velocity dispersion 
anisotropy with respect to the density slope, $\gamma_{fit}$.  This relation 
has been presented for N-Body simulations by \cite{hm04}. They have found that 
there is a tight correlation between $\beta$ and $\gamma$ irrespective of the 
considered models.  ESIM halos do not display this correlation. One possible 
explanation for this difference between ESIM and N-Body halos is that ESIM 
formalism does not allow the onset of the radial orbit instability, which may 
be key in forming $\beta-\gamma$ correlation in N-Body halos \citep{barnes05a}.

\subsubsection{Phase-space density}\label{psd}

\begin{figure}[t]
\plotone{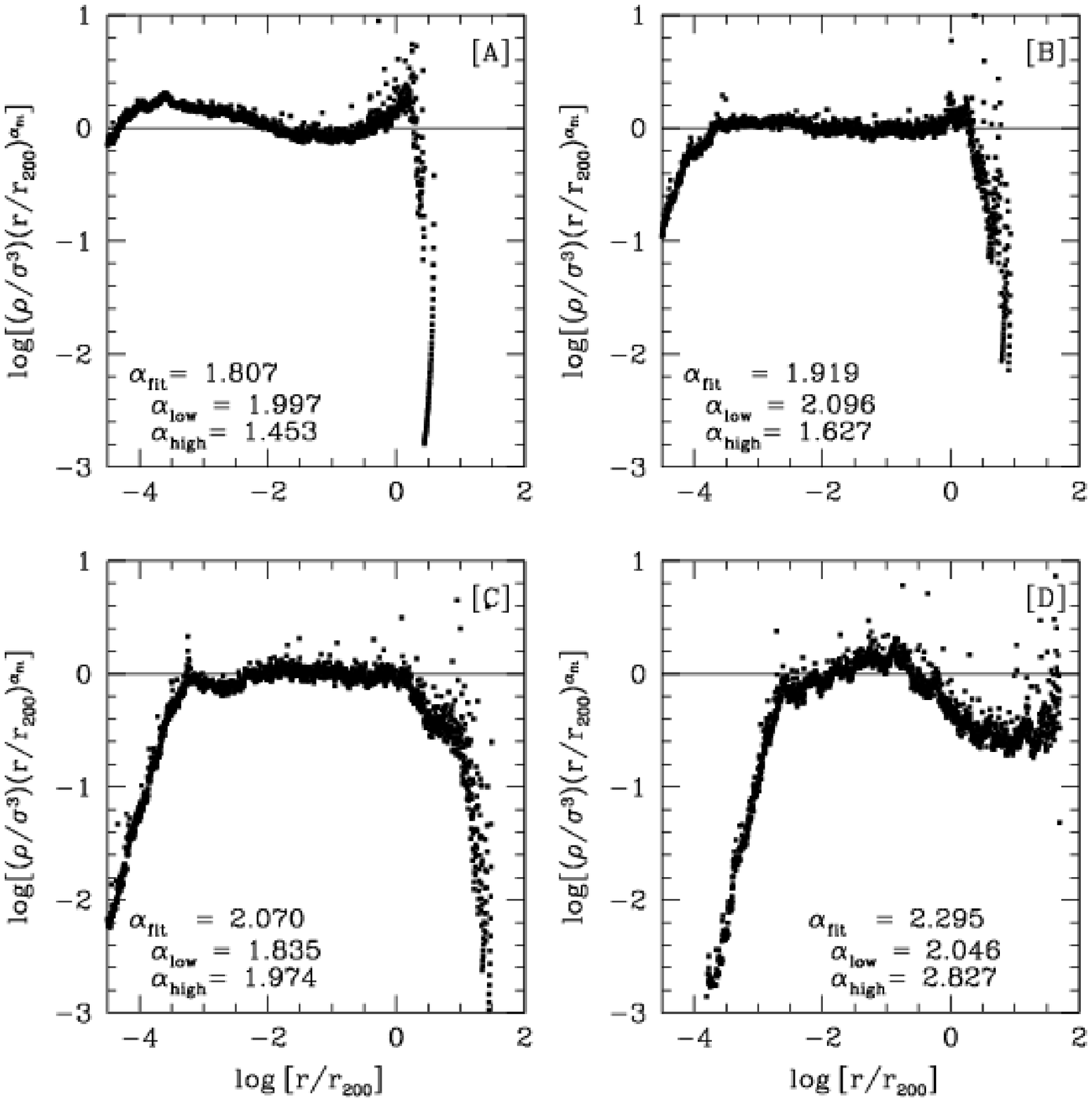}
\caption{
	The phase-space density profiles of halos with perturbation amplitude
        \factos\/ of 0.4 [A], 0.6 [B], 1 [C], and 2 [D]. The \powerlaw\/ data 
        have been normalized to a power-law fit between $10^{-2.5}$ and 1 virial 
        radius.  The value of the power-law index fit is indicated in each panel
        as $\alpha_{fit}$.  Also, the slope of the \powerlaw\/ profiles from 
        $r/r_{200}$ of $[10^{-2.5}, 10^{-2}]$ and $[10^{-0.5}, 1]$ are 
        noted as $\alpha_{low}$ and $\alpha_{high}$ for comparison.  Note the 
        nearly scale-free behavior for 2-3 decades in radius.
	} 
\label{powerlaw}
\end{figure}

The phase-space density profiles of the halos displayed in Figure~\ref{density}
are shown in Figure~\ref{powerlaw}. ESIM halos appear to have nearly 
scale-free distribution of \powerlaw\/ extending over $2-4$ decades in radius,
interior to the virial radius. Aside from the very central and outer regions 
(which we will return to shortly) \powerlaw\/ are fit by power-laws better than 
the corresponding density profiles. For example, between $10^{-2.5}$ and $10^{-3.5}$ 
virial radii the density profile of the halo with the \facto\/ of 1 (panel [C]) is 
becoming shallow rapidly, but \powerlaw\/ is still well fit by the 
same slope that characterizes larger radii.

\begin{figure}[t]
\plotone{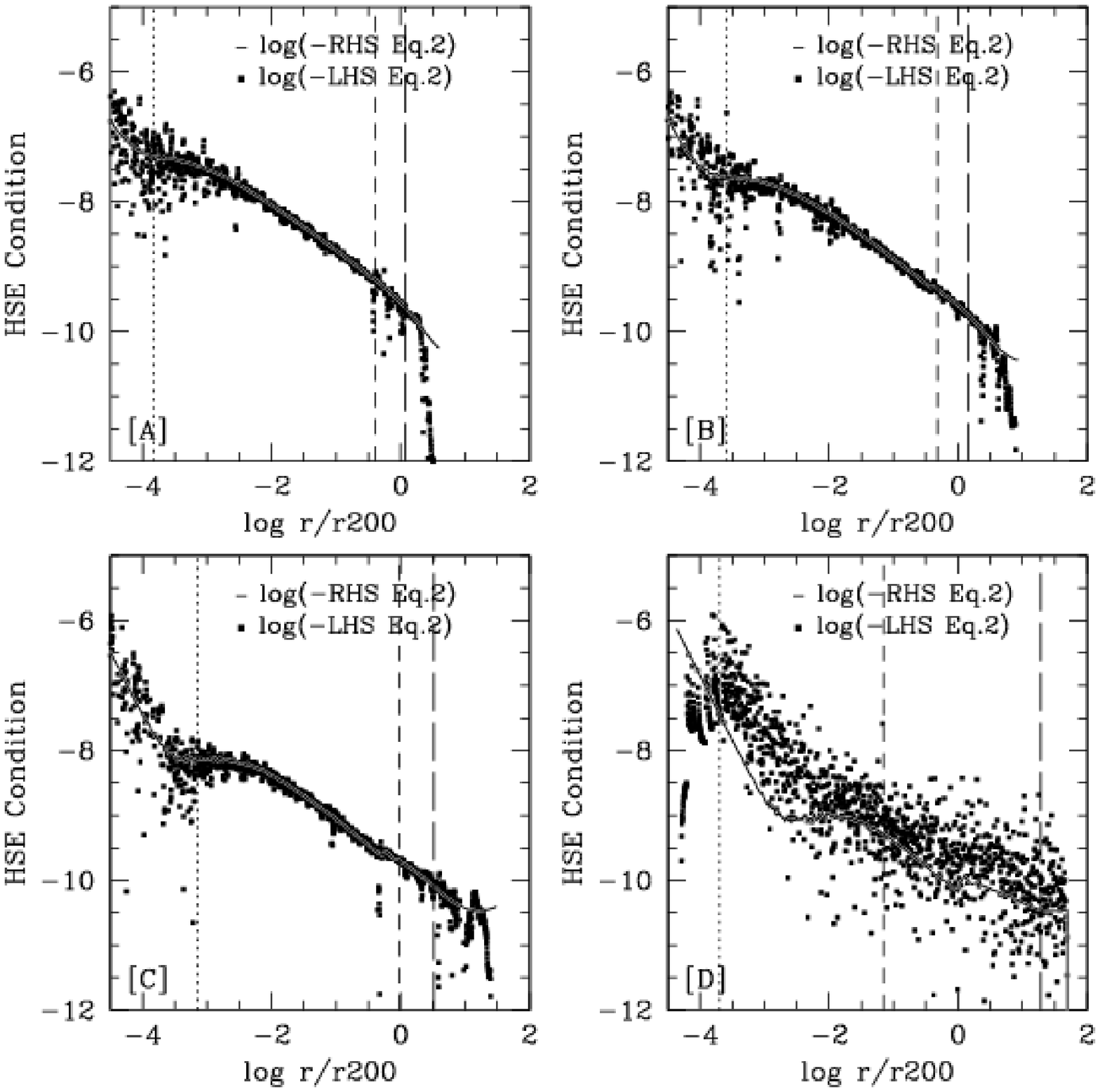}
\caption{The hydrostatic equilibrium (HSE) condition: the LHS and RHS of equation 
      \ref{hse_eq} are shown as points and the smooth curve, respectively.
      Panels A though D represent halos with \factos\/ of
      0.4, 0.6, 1, and 2, respectively.  Where the line traces the plotted 
      points, the halo is in hydrostatic equilibrium. The dotted, short dash 
      and long dash lines vertical lines are the same as in Figure~\ref{density}.
      } 
\label{hse_fig}
\end{figure}

To understand why \powerlaw\/ deviates from a power-law at large and small
radii we have to consider the equilibrium state of halos. \citet{BT}, 
under the assumptions of spherical symmetry and a non-rotating system, 
rewrite the Jeans equation as 
\begin{equation}
{\frac{1}{\rho}\frac{d(\rho\sigma^2_r)}{dr}+2\frac{\beta\sigma^2_r}{r}=-\frac{d\phi}{dr}.}
\label{hse_eq}
\end{equation}
In Figure~\ref{hse_fig}, the LHS of equation~\ref{hse_eq}, i.e. the pressure
gradient plus the anisotropy term, is plotted at each radius as points, 
while the RHS, the potential gradient, is shown as the smooth curve.
Hydrostatic equilibrium is satisfied when the points follow the curve.  
This is the case in panels [A], [B], and [C]; in panel [D] there is a lot of
scatter in the points that represent the LHS of eq.~\ref{hse_eq}, and so
the equilibrium condition is not well satisfied. This apparently results in
\powerlaw\/ being only marginally well approximated by a power-law. In
panels [A]--[C], the hydrostatic equilibrium breaks down outside of the
virial radius, which apparently leads to a break in the scale-free behavior
of \powerlaw, at the same radius.

Figure~\ref{hse_fig} also helps to explain why \powerlaw\/ breaks away from
a power-law at small radii. At small radii, the potential gradient scales as
$1/r^2$. This is due to the presence of a compact inert core at the center
of every ESIM halo. As described in ~\citet{wbd04} (Sections 4.1, A.1, A.5),
ESIM halos have finite mass resolution. At the center of each halo the smallest 
resolvable mass is represented by an inert core, which provides the initial 
`seed' for ESIM calculations. The inert core does not evolve, i.e. its mass 
distribution does not change as the halo collapses. Because the part of the 
halo where the core dominates is not subject to the same dynamical evolution 
as the rest of the halo, that part should not be considered in the analysis. 
In panels [A]--[D] of Figure~\ref{hse_fig} the regime of core domination 
begins roughly at $10^{-4}$, $10^{-3.5}$, $10^{-3}$, and $10^{-2.5}$ virial 
radii, respectively. The corresponding panels of Figure~\ref{powerlaw} show 
that these radii coincide with the break in the scale-free nature of \powerlaw. 

Excluding the innermost and the outermost regions of halos, the remaining 
portions are well represented by power-laws (this is further illustrated in 
Section~\ref{rgvalues} and Figure~\ref{rg035070} below). Figure~\ref{powerlaw} 
demonstrates that the scale-free nature of \powerlaw\/ is resilient: it persists 
despite the changes in the magnitude of the secondary perturbations, at least
in the range 0.5-1.5. However, the power-law index, 
$\alpha=-(d\ln\rho/\sigma^3)/(d\ln r)$, is not constant; in general, 
$\alpha_{fit}$ increases with increasing \facto\/ 
(see panel B of Figure~\ref{vsfacto}).

\subsubsection{Halo mass}\label{mass}

Panel D of Figure \ref{vsfacto} shows that the mass of virialized halos, for a 
given power spectrum filtering length ($r_h=0.05$Mpc for these halos), decreases 
steadily as the perturbation amplitude \facto\/ increases.  Large secondary 
perturbation amplitude implies a more extended mass distribution, and smaller 
virial masses.  Additionally, if the shells' energy reaches the threshold to 
cause mass loss, the final mass is reduced further. However,
this effect is negligible for most halos: there is no mass loss 
during formation for halos with the \facto\/ less than about 1.5.

\subsubsection{ESIM halos of galaxy groups and cluster masses.}\label{rgvalues}

\begin{figure}[t]
\plotone{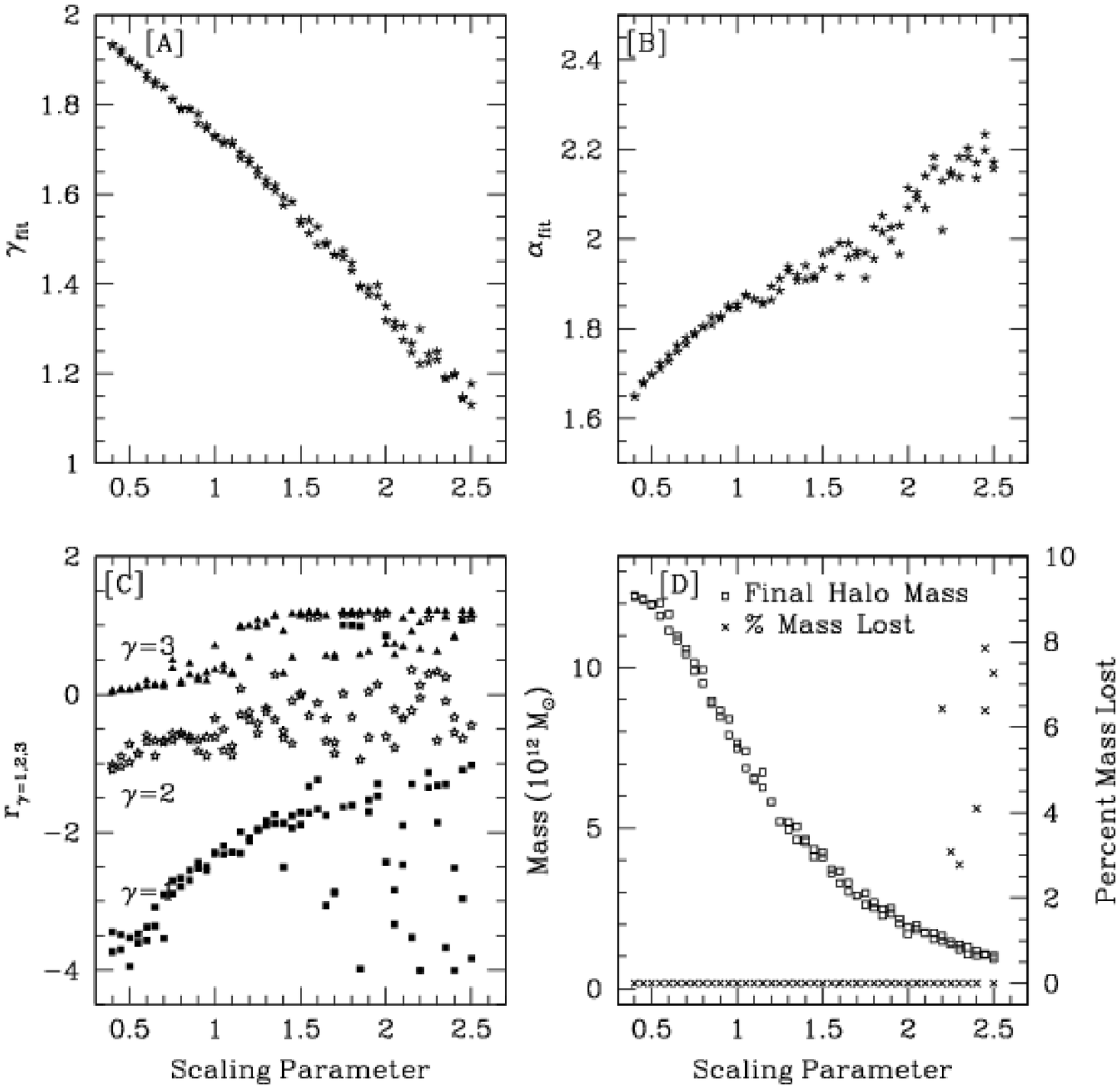}
\caption{Same as Figure~\ref{vsfacto}, but for group-mass halos with $r_h=0.35$Mpc.}
\label{rgvsfacto}
\end{figure}

\begin{figure}[t]
\plotone{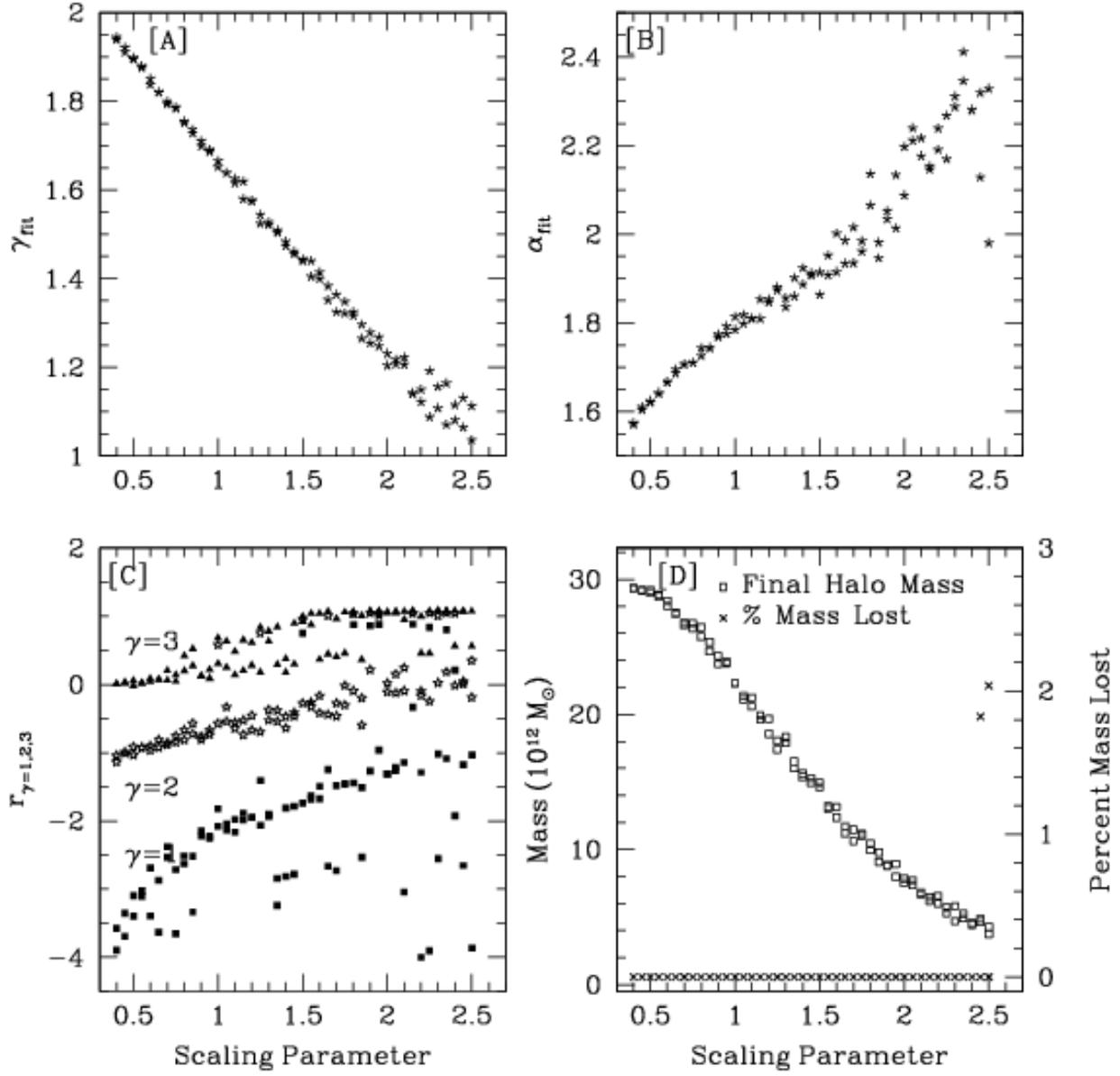}
\caption{Same as Figure~\ref{vsfacto}, but for cluster-mass halos with $r_h=0.7$Mpc.}
\label{rg2vsfacto}
\end{figure}

So far, we have discussed how the properties of galaxy-mass halos ($r_h=0.05$Mpc) 
change with the changing amplitude of secondary perturbations. Group and cluster-mass 
halos, generated using power spectrum filtering lengths of $r_h=0.35$ and 0.7Mpc, 
respectively, behave similarly. The overall trends for these two cases are 
summarized in Figures~\ref{rgvsfacto} and \ref{rg2vsfacto}, which are similar to 
Figure~\ref{vsfacto}.  

\begin{figure}[t]
\plotone{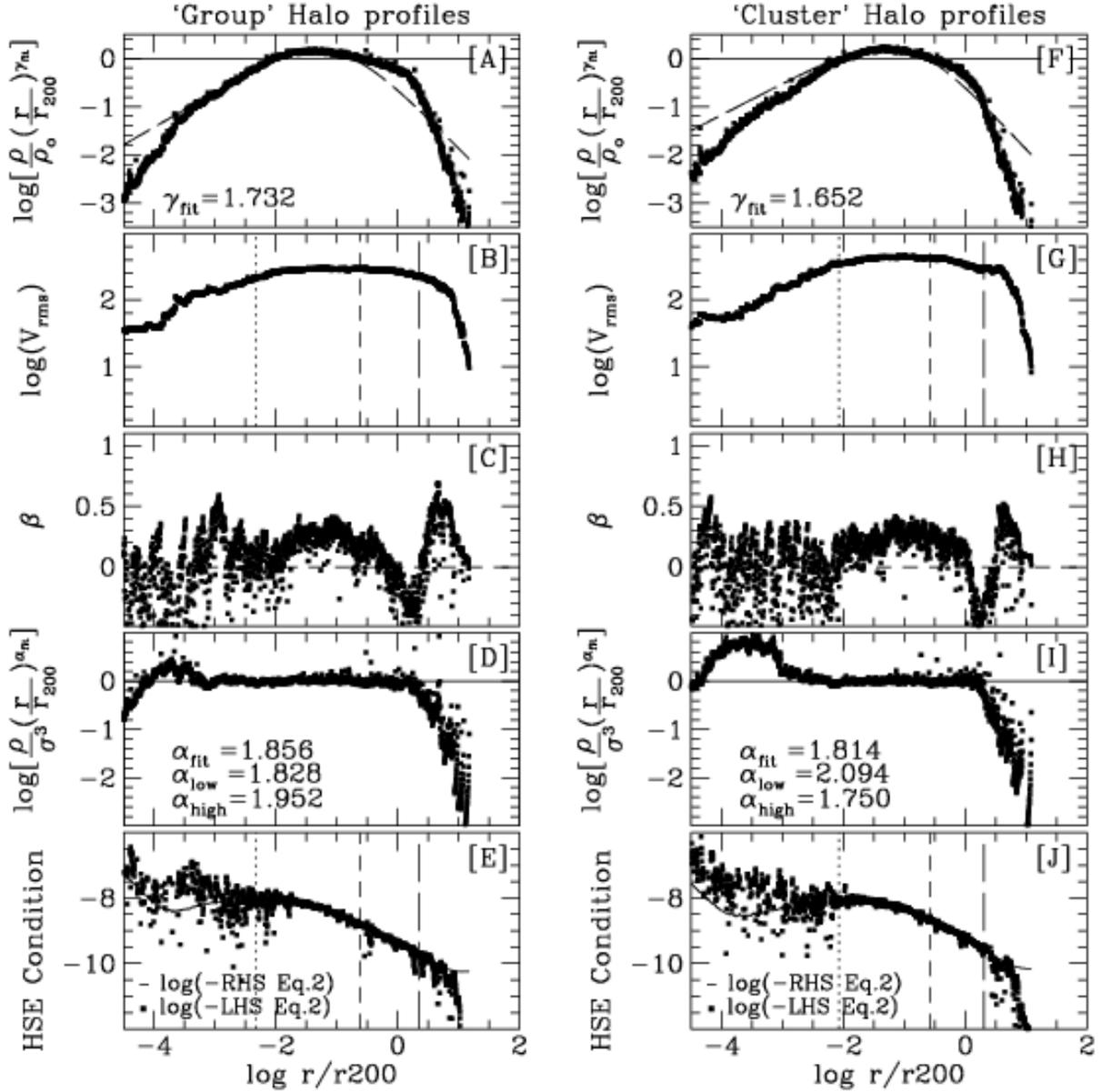}
\caption{
  	 Profiles of two halos, both having \facto\/ of 1, generated with $r_h=0.35$Mpc 
        (panels A through E) and $r_h=0.7$Mpc (panels F through J). Shown, from top to
        bottom, are the density profiles, with the power-law best fit taken out, as in
        Figure~\ref{density}
        (the NFW formula is over-plotted as a long dash line), the RMS velocity 
        dispersion, the velocity anisotropy $\beta$, the phase-space density with the
        power-law best fit taken out, and the equilibrium condition, showing the LHS 
        (points) and the RHS (smooth curve) of eq.~\ref{hse_eq}.
        Again, the values of $\gamma_{fit}$, $\alpha_{fit}$, $\alpha_{low}$, and
        $\alpha_{high}$ are noted in the appropriate panels.
        Note that the phase-space density profiles are well fit by power-laws, 
        while the density profiles are not.
        } 
\label{rg035070}
\end{figure}

Two examples are shown in Figure~\ref{rg035070}, for $r_h=0.35$Mpc (panels A 
through E) and $r_h=0.7$Mpc (panels F through J); both halos have \facto\/ of 1. 
It is apparent that  larger $r_h$ values result in density profiles that are 
not as well described by a single power-law, as smaller $r_h$ halos. This is 
further illustrated in panels C of Figure~\ref{rgvsfacto} and \ref{rg2vsfacto}; 
$r_{\gamma=1}$, $r_{\gamma=2}$ and $r_{\gamma=3}$ are closer together than in 
galaxy-mass halos, implying that the density slope changes more rapidly with radius.

Despite the non power-law behavior of density, the phase-space density,
\powerlaw, remains a power-law for at least 2-3 decades in radius interior 
to $r_{200}$ (panels D and I of Figure~\ref{rg035070}). These panels also
show that the scale-free nature of \powerlaw\/ breaks down interior to 
$10^{-3.5}-10^{-3}$, where the inert core begins to dominate 
the potential (panels E and J). The same behaviour was noted in the case of
galaxy-mass halos, in Section~\ref{psd}. The best-fit 
slope $\alpha_{fit}$ increases with the \facto\/ for a fixed $r_h$ 
(panels B of Figure~\ref{vsfacto}-\ref{rg2vsfacto}), but gets consistently 
smaller with increasing $r_h$. 

\begin{figure}[t]
\begin{center}
\plotone{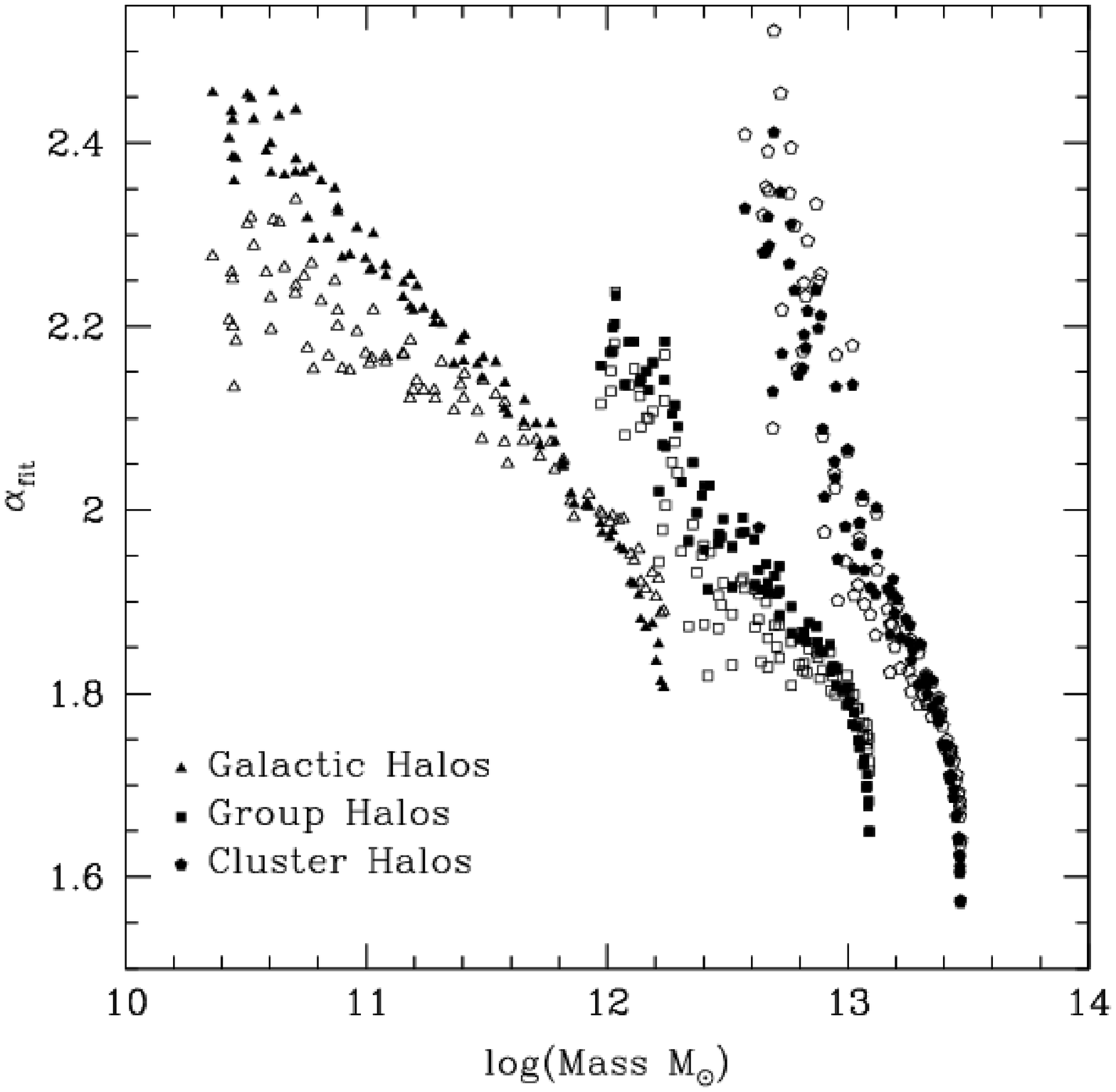}
\caption{
	The double logarithmic slope, $\alpha_{fit}$, of \powerlaw\/ vs. radius,
        as a function of the virial halo mass, for three different $P(k)$ filtering
        scales, $r_h=0.05, 0.35$ and 0.7Mpc (triangles, squares, and pentagons).
        Filled (empty) symbols show $\alpha_{fit}$ values computed using 
        $10^{-2.5}\rightarrow 1$ ($10^{-2.5}\rightarrow 10^{-0.5}$) radial range,
        given in terms of $r_{200}$. Note that $\alpha_{fit}$ is not uniquely
        determined by halo mass, amplitude of secondary perturbations, halo
        filtering scale, etc.
}
\label{m_vs_a}
\end{center}
\end{figure}

Figure~\ref{m_vs_a} depicts the dependence of the \powerlaw\/ power-law 
slope, $\alpha_{fit}$ on the virial mass for all three sets of halos: $r_h=0.05$ Mpc 
(triangles), $r_h=0.35$ Mpc (squares), and $r_h=0.7$ Mpc (pentagons). 
Filled symbols show $\alpha_{fit}$ obtained by fitting a power-law to the 
data within our standard radial range, from $10^{-2.5}$ to 1 virial radii. 
To test the dependence of the specific choice of the radial range we also 
use a reduced range of $10^{-2.5}$ to $10^{-0.5}$ virial radii (empty symbols). 
The $\alpha_{fit}$ values are similar in both cases. The main conclusion from 
Figure~\ref{m_vs_a} is that $\alpha_{fit}$ is not uniquely determined 
by any single halo property, such as mass, 
perturbation amplitude \facto, or $r_h$.

\subsection{Testing the Resiliency of the \powerlaw\/ Power-law}
\label{resiliency}

Since the halos considered in previous sections exhibit a near-perfect 
power-law in \powerlaw, we wish to determine what would be required to break this
resilient behavior.  To that end, we generate halos with different \factos\/
assigned to blocks of shells.  These changes are not meant to mimic realistic 
situations, but rather to push our simulations to unphysical extremes. In several 
trials we increased or decreased the \facto\/ linearly with radius; in other 
trials, we assigned the inner 5-15\% of radii a larger perturbation amplitude
\facto\/ of 5, 20, or 50, with smaller \factos\/ assigned to outer shells.

\begin{figure}[t]
\plotone{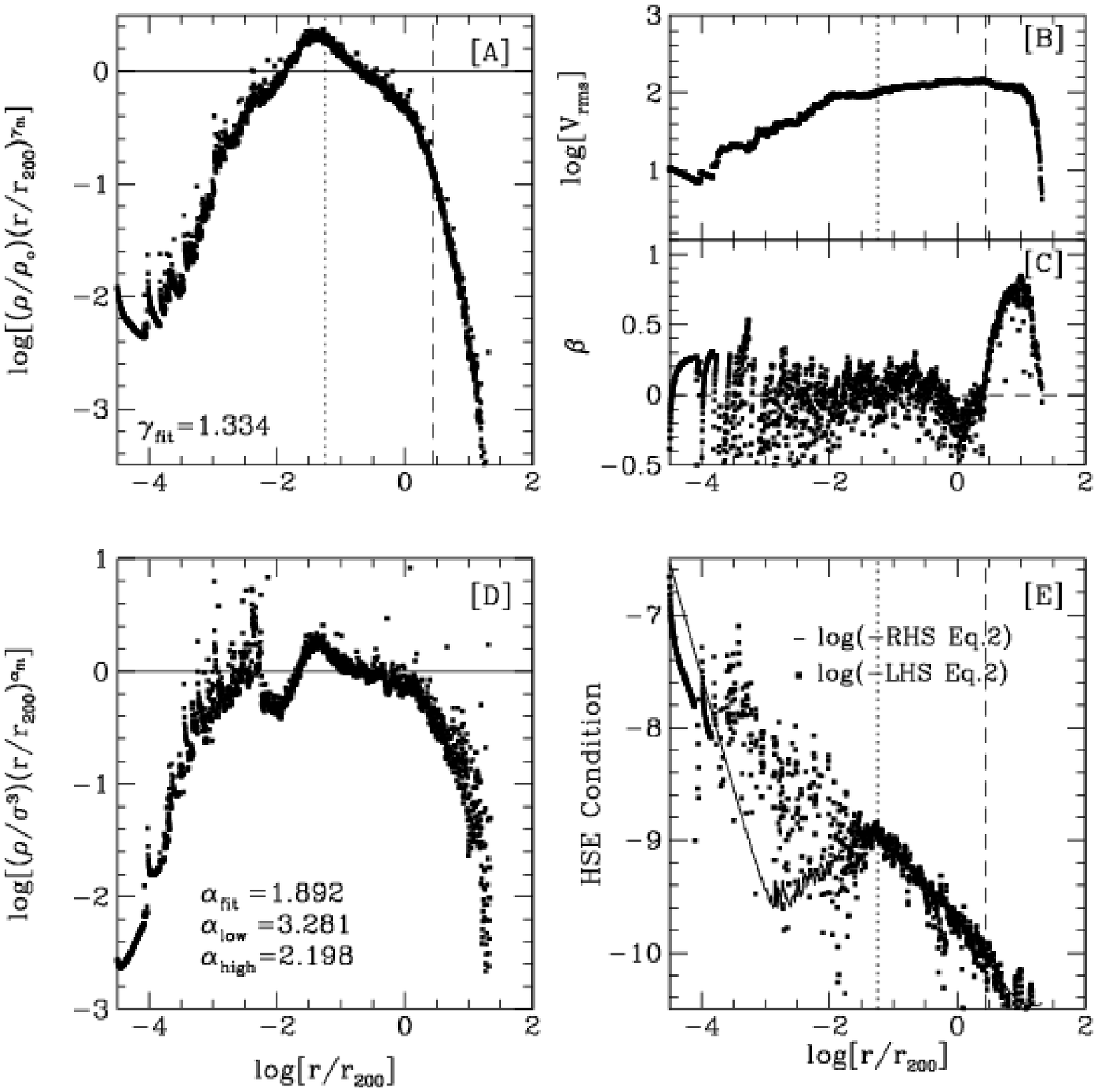}
\caption{
	An example of a halo with discontinuous perturbation amplitude \factos: 
        shells with the 
        initial comoving radii between $0.59$ and $0.79$ Mpc were assigned a 
        \facto\/ of 5, while the rest of the halo (up to the last 
        shell at 1.92 Mpc) were assigned a \facto\/ of 0.9. 
	[A]: Density profile, normalized to a power-law fit between 
           $r/r_{200}$ of $10^{-2.5}$ and $1$;
	[B]: RMS velocity dispersion profile in $km/s$;
	[C]: Velocity dispersion anisotropy profile, $\beta$. 
      [D]: Phase-space density profile, normalized to its power-law fit;
	[E]: The equilibrium condition of equation~\ref{hse_eq}: the dots 
           represent the LHS and the thin line the RHS.  Where the line traces
           the plotted points, the halo is in hydrostatic equilibrium.
	} 
\label{brokenlaw}
\end{figure}

The halos with the \facto\/ assigned linearly in radius (with no large 
discontinuities) display the power-law in \powerlaw.  However, halos in the 
trials with the `excited sections' (blocks of shells with large \factos) had 
a \powerlaw\/ profile that showed deviations from a power-law behavior. 
Figure~\ref{brokenlaw} shows the specific halo where all the shells were 
assigned a \facto\/ of 0.9, except for those with an initial comoving radius 
between $0.59$ and $0.79$ Mpc, which were assigned a \facto\/ of 5.  The last 
shell to collapse in this evolution had a comoving radius of 1.92 Mpc.  
Visual inspection of panels A and B show that, even though these halos have
been subjected to radially discontinuous changes in the amplitude of the 
secondary perturbations, the final density and velocity dispersion profiles
still appear well behaved,
but the deviations of \powerlaw\/ from a power-law are evident.
This and other similar trials illustrate that one must adopt
extreme measures---large discontinuities in the \facto---in order to break 
the scale-free nature of the phase-space density profile.

\subsection{Conclusions from ESIM Halos}\label{ESIMconc}

The ESIM formalism provides us with a simplified, but physically 
self-consistent way of generating collisionless dark matter halos. It is very 
different from N-body simulations. In the preceding sections
we described ESIM halo properties in detail, pointing out that the density 
profiles are generally different from those produced in N-body simulations. 
By changing the initial conditions we produced a range of density profile 
types: by filtering the mass power spectrum on small scales ($r_h=0.05$Mpc) 
to simulate galaxy-mass halo formation, we ended up with 
density profiles that showed virtually no
change in slope over 2-3 decades in radius inside of the virial radius
(Figure~\ref{density}). Larger filtering scales ($r_h=0.35$ and 0.7 Mpc) 
aimed to produce group and cluster-mass halos resulted in more curved
density profiles (Figure~\ref{rg035070}). 

The most important conclusion of Section~\ref{ESIMsec} concerns the
phase-space density profile, \powerlaw.  Despite the differences in methods 
and density profile shapes, one feature was shared by all ESIM halos and 
N-body simulated halos: the power-law nature of the \powerlaw\/ profile. 
We have shown that this feature of ESIM halos is very resilient, and one 
must concoct rather extreme cases in order to end up with \powerlaw\/ 
that shows deviations from a single power-law (Section~\ref{resiliency}). 
There are no models in astrophysics literature that provide a physical explanation 
for this feature of virialized halos. We argue that the scale-free nature
\powerlaw\/ cannot be uniquely characteristic of hierarchical merging 
because ESIM halos do not undergo mergers. We speculate that it is a 
generic consequence of the one dynamical process which is common to both 
ESIM and N-body halo formation---violent relaxation, or, global changes 
in the potential, and the accompanying changes in particle energies 
brought about by the collapse.

Our results concerning $\alpha$, the double logarithmic slope of 
\powerlaw\/ vs. radius, are different from those of TN01. We find a range 
of values from 1.6 to 2.2, while TN01 find that the same $\alpha=1.875$ 
described all three of their N-body halos. It is possible that the latter 
finding is due to a small number of halos studied. In our study, $\alpha_{fit}$ 
was not uniquely determined by any single property of halos, such as mass, 
epoch of formation, or amplitude of random velocities.

\section{Jeans equation analysis}\label{jeans}

\subsection{The second derivative of the Jeans equation}\label{eqn6}

The findings of the preceding section, summarized in 
Section~\ref{ESIMconc}, motivate our next step. ESIM halos are 
spherically symmetric, in equilibrium, have zero net angular momentum,
and have approximately isotropic velocity dispersion ellipsoids, 
$\beta\approx 0$.\footnote{Looking at Figure~\ref{velocity} 
zero anisotropy is arguably the least justified of our assumptions.
N-body generated halos have velocity dispersions that are nearly 
isotropic at the center, and become radially anisotropic further out. 
Strictly speaking, the analysis in this Section applies only to the 
portions of halos that are isotropic. We will extend our analysis
to arbitrary anisotropy in the future work. As a place to start, 
$\beta\approx 0$ is a reasonable approximation.} 
These halo properties lead directly to the appropriate
version of the Jeans equation, and its first derivative:
\begin{equation}
{{d(\rho\sigma^2)}\over{dr}}=-\rho{{GM(<r)}\over{r^2}},
\label{tn01eq1}
\end{equation}
which states that halos are supported against collapse by ``pressure,''
provided by the velocity dispersion of dark matter particles.  Because
halos are assumed to have isotropic velocity dispersions,
$\sigma=\sigma_r=\sigma_\theta=\sigma_\phi$.
In Section~\ref{ESIMsec} we have established empirically that ESIM halos 
exhibit scale-free \powerlaw\/ behavior. In this paper we will not address 
the question of {\em why\/} phase-space density radial profiles are 
power-laws. Instead, we will take this property as an additional constraint:
\begin{equation}
{\rho\over\sigma^3}(r)= {\rho_0\over\sigma_0^3}\Bigl({r\over
  r_0}\Bigr)^{-\alpha}.
\end{equation}
where $\rho_0$, $r_0$, $\sigma_0$, and $\alpha$ are constants for any given 
halo. Eliminating $\sigma$, and using the dimensionless variables, $x=r/r_0$ and 
$y=\rho/\rho_0$, we can now write down the derivative of equation~\ref{tn01eq1},
\begin{equation}
{d\over{dx}}\Bigl[{{-x^2}\over{y}}{d\over{dx}}(y^{5/3}x^{2\alpha/3})\Bigr]
=\kappa y x^2.
\label{tn01eq4}
\end{equation}
This is also equation~4 of TN01. The constant $\kappa=4\pi G\rho_0 r_0^2/\sigma_0^2$ 
specifies the normalization of halos. Exponent $\alpha$ is constant for any 
given halo, but we leave its value unspecified. 

In the rest of this Section we develop a framework for studying halos that can 
be described by equation~\ref{tn01eq4}. However, unlike the earlier Jeans 
equation work our main equation will not be equation~\ref{tn01eq4} itself, but 
its first derivative, which is also the second derivative of the Jeans equation:
\begin{equation}
(2\alpha+\gamma-6)(2\alpha-5\gamma)(\tth[\alpha-\gamma]+1)=
3\gamma\p(8\alpha-5\gamma-5)+15\gamma\pp,
\footnote{This equation corrects a sign error of eq.(2.2) of \cite{confproc}.
Note that equation~\ref{b1}, together with some interesting analysis also 
appears in \cite{dm05}.}\label{b1}
\end{equation}
where we have defined logarithmic derivatives of the density slope, 
$\gamma=-d\ln y/d\ln x$, as $\gamma\p=d\gamma/d\ln x$, and 
$\gamma\pp=d\gamma\p/d\ln x$. Equation~\ref{b1} has several advantages over 
equation~\ref{tn01eq4}. It does not depend on the normalization constant 
$\kappa$, and has no explicit dependence on $x$, in fact, equation~\ref{b1} 
depends only on the exponent $\alpha$, and the local density slope, $\gamma(x)$. 
Equation~\ref{b1} looks like an algebraic equation---its partially factorized 
form will prove useful in exploring the various types of solutions of the 
Jeans equation.

Our goal in the following is to explore solutions of the Jeans equation
in general, paying particular attention to the role of $\alpha$. 
Specifically, we want to know if the Jeans equation admits any 
special $\alpha$ values.

\subsection{The $\alpha-\gamma$ plane}\label{alphabeta}

\begin{figure}[t]
\epsscale{0.7}
\plotone{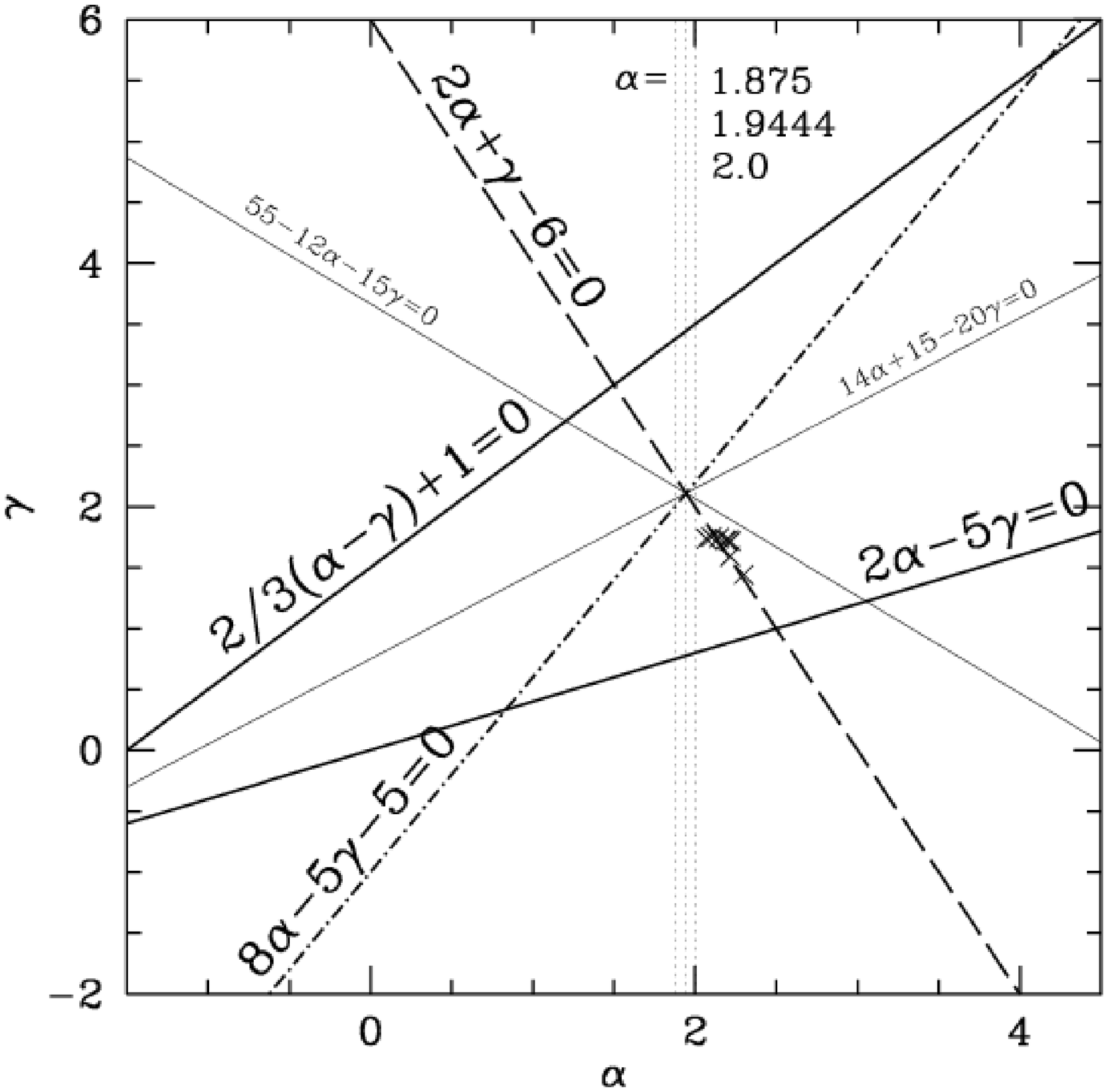}
\caption{Pictorial depiction of equation~\ref{b1}. Thick solid and long-dash 
lines represent the terms on the LHS of equation~\ref{b1}. The dot-dash line
is the coefficient of the damping term on the RHS of equation~\ref{b1}.
The two thin solid lines are $\gamma_0(\alpha)$'s given in equations~\ref{mf} 
and \ref{convexmf}. The three vertical dotted lines at $\alpha=1.875, 1.9444$ 
and $2$ are plotted to guide the eye. Periodic solutions are obtained for 
$\alpha=1.9444$ (Section~\ref{periodic}). A set of closely spaced crosses are 
$\alpha_{fit}$ and $\gamma_{fit}$ values for galaxy-mass ESIM halos with 
$\beta\approx 0$.}
\label{smtriangle}
\end{figure}

A halo is characterized by a relation between $\gamma$ and $x$, which can 
be obtained by assuming a value for $\alpha$, specifying initial conditions 
for $\gamma$ and $\gamma\p$, and then integrating equation~\ref{b1}. An
infinite 
variety of halos can be generated in this fashion, so a straightforward
exploration of the parameter space is probably not the most insightful one.
The form of equation~\ref{b1}, namely that it looks like a factorized algebraic
equation of two variables, $\alpha$ and $\gamma$, suggests that 
it is instructive to consider the $\alpha-\gamma$ plane, shown in 
Figure~\ref{smtriangle}. Assuming $\alpha$ is constant for any given 
halo, a halo density profile can be represented by a vertical straight 
line segment. 

There are four factors in equation~\ref{b1}, three on the LHS, and one on 
the RHS. By separately setting these to zero we obtain the four thick lines 
in the $\alpha-\gamma$ plane. This plot provides a framework for understanding 
and classifying different types of solutions of the Jeans equation.

Consider the line $2\alpha-5\gamma=0$ (lower thick solid line). If part 
of a halo is described by a constant $\gamma=(2/5)\alpha$ for some range 
of radii, then the LHS of equation~\ref{tn01eq1} can be expressed as 
${{d\ln(\rho\sigma^2)}/{d\ln r}}\propto 2\alpha-5\gamma$, implying
that the pressure gradient vanishes in that part of the halo.
An example of this is the central region of the critical solution 
identified by TN01, where $\gamma=0.75$, corresponding to $\alpha=1.875$.
In general, halos of any $\alpha$ whose inner profile slope is described
by $2\alpha-5\gamma=0$ have vanishing pressure gradient in the center.

The role of terms $2\alpha+\gamma-6$ and $8\alpha-5\gamma-5$ is addressed 
in Sections~\ref{betapl} and \ref{periodic} respectively. The term 
$\tth[\alpha-\gamma]+1$ is briefly discussed in Section~\ref{convex}, and
all four terms of equation~\ref{b1} are important in the discussion of 
Section~\ref{periodic}.

We note that halos with radially increasing density cannot exist;
\cite{tremaine94} showed that 
if the distribution function depends solely on energy, as is the
case with our isotropic halos, then $\gamma<0$ is unphysical.

\subsection{Power-law density profiles}\label{betapl}

One of the factors on the LHS of equation~\ref{b1} is $2\alpha+\gamma-6$;  
$2\alpha+\gamma-6=0$ is shown as the long-dash line in Figure~\ref{smtriangle}.  
$\gamma_{pl}=6-2\alpha$ represent power-law density profile solutions to the 
Jeans equation (TN01, \cite{h04}), i.e. if 
$\gamma\p=\gamma\pp=0$ then $\gamma_{pl}$=const for all $x$. 
A power-law density profile is represented by a point in Figure~\ref{smtriangle}. 
For example, the singular Isothermal Sphere is a viable solution of 
equation~\ref{b1}, characterized by $\alpha=\gamma=2$ 
and constant $\sigma$ throughout the halo. 

Because equation~\ref{b1} does not distinguish between the three factors
on the LHS, one might think that each of these three should produce
power-law density profiles. This is not so. Equation~\ref{tn01eq4} can be 
rewritten as
$(2\alpha-5\gamma)(\tth[\alpha-\gamma]+1)-5\gamma\p=-3\kappa\;x^{2-\tth\alpha}\;y^{1/3}$.
If the slope $\gamma$ is given by $2\alpha-5\gamma=0$, or $\tth[\alpha-\gamma]+1=0$, 
and $\gamma\p=0$ because we are dealing with a power-law density profile,
then the RHS becomes zero, which is unphysical. (Recall that $y$ is scaled density.) 
Therefore, for a given $\alpha$ there is a unique power-law density solution, 
$\gamma_{pl}=6-2\alpha$.

The one-to-one correspondence between $\gamma_{pl}$ and $\alpha$ can be 
tested using ESIM halos. Galaxy-mass ($r_h=0.05$ Mpc) ESIM halos presented 
in Section~\ref{ESIMhalos} have nearly power-law density profiles. Of these, 
halos with the perturbation amplitude \factos\/ of 1 or larger have nearly 
isotropic velocity dispersion ellipsoids. For these halos the relation between 
the fitted slopes of the density and phase-space density profiles should follow 
$2\alpha_{fit}+\gamma_{fit}-6=0$.  In Figure~\ref{smtriangle} we plot 
$\alpha_{fit}$ and $\gamma_{fit}$ for halos with \factos\/ between 1 and 
2.5 (crosses). There is good agreement between ESIM halo data and predictions.

\subsection{The unique periodic solution of Equation~\ref{b1}}\label{periodic}

Equation~\ref{b1} is a non-linear damped oscillator equation for $\gamma$,
with the $\gamma\p$ term being analogous to a dissipation term.
Equation~\ref{b1} therefore suggests that the Jean's equation may admit
solutions where the slope of the density profile oscillates with radius, 
becoming alternately steeper and shallower.  If equation~\ref{b1}
is in a regime where these oscillations are damped, then the slope
$\gamma$ may oscillate, with the amplitude of
the oscillations decaying with increasing radii. Asymptotically, this 
behavior would lead to a density profile that has a power-law fall-off 
at large radii. Instead, if equation~\ref{b1} is in a regime where the 
oscillations can amplify, then the slope may run away to large values, 
leading to a density profile that becomes increasingly steep at large radii.  

The coefficient of the $\gamma\p$ term in equation~\ref{b1} controls the 
``dissipation'' in the oscillations of the density profile slope $\gamma$. If 
this coefficient were always positive (negative),  it would damp (amplify) the 
amplitude of $\gamma$ oscillations with $\ln x$.\footnote{We stress that the 
halos are in equilibrium, and the use of the word ``oscillations'' does not 
mean that the halos are changing in time: the changes are in the density 
profile slope, as a function of logarithmic radius.}  If the coefficient is
zero, the amplitude of oscillations would not evolve, resulting in perpetually 
periodic oscillations. In the case of equation~\ref{b1}, the damping term is 
non-linear, with its sign changing as a function of $\ln x$. Does a periodic 
solution exist in such a situation? 

Let us use Figure~\ref{smtriangle} as a guide. Notice that the point of 
intersection of $8\alpha-5\gamma-5=0$ and $2\alpha+\gamma-6=0$ lines is 
exactly half-way between $2\alpha-5\gamma=0$ and $\tth[\alpha-\gamma]+1=0$. 
The corresponding $\alpha=35/18=1.9444$, and with that equation~\ref{b1} can 
be rewritten as
\begin{equation}
\gamma\pp=\frac{2}{9}B\Bigl(\Bigl[\frac{4}{3}\Bigr]^2-B^2-\frac{9}{2}\gamma\p\Bigr),
\label{b1periodic}
\end{equation}
where $B=\gamma_{pl}-\gamma$, and $\gamma_{pl}=2.111$ for $\alpha=1.9444$.
Equation~\ref{b1periodic} says that when $B=0$, then $\gamma\pp=0$,
and hence $\gamma\p$ attains a local minimum or maximum. So, every time 
$\gamma=2.111$, $\gamma\p$ attains an extremum. This statement can be true 
only if $\gamma$ varies periodically with $\ln x$. Getting back to the
coefficient of the damping term: when $\alpha=1.9444$, the value of
$8\alpha-5\gamma-5$ is, on average, zero.

\begin{figure}[t]
\epsscale{0.6}
\plotone{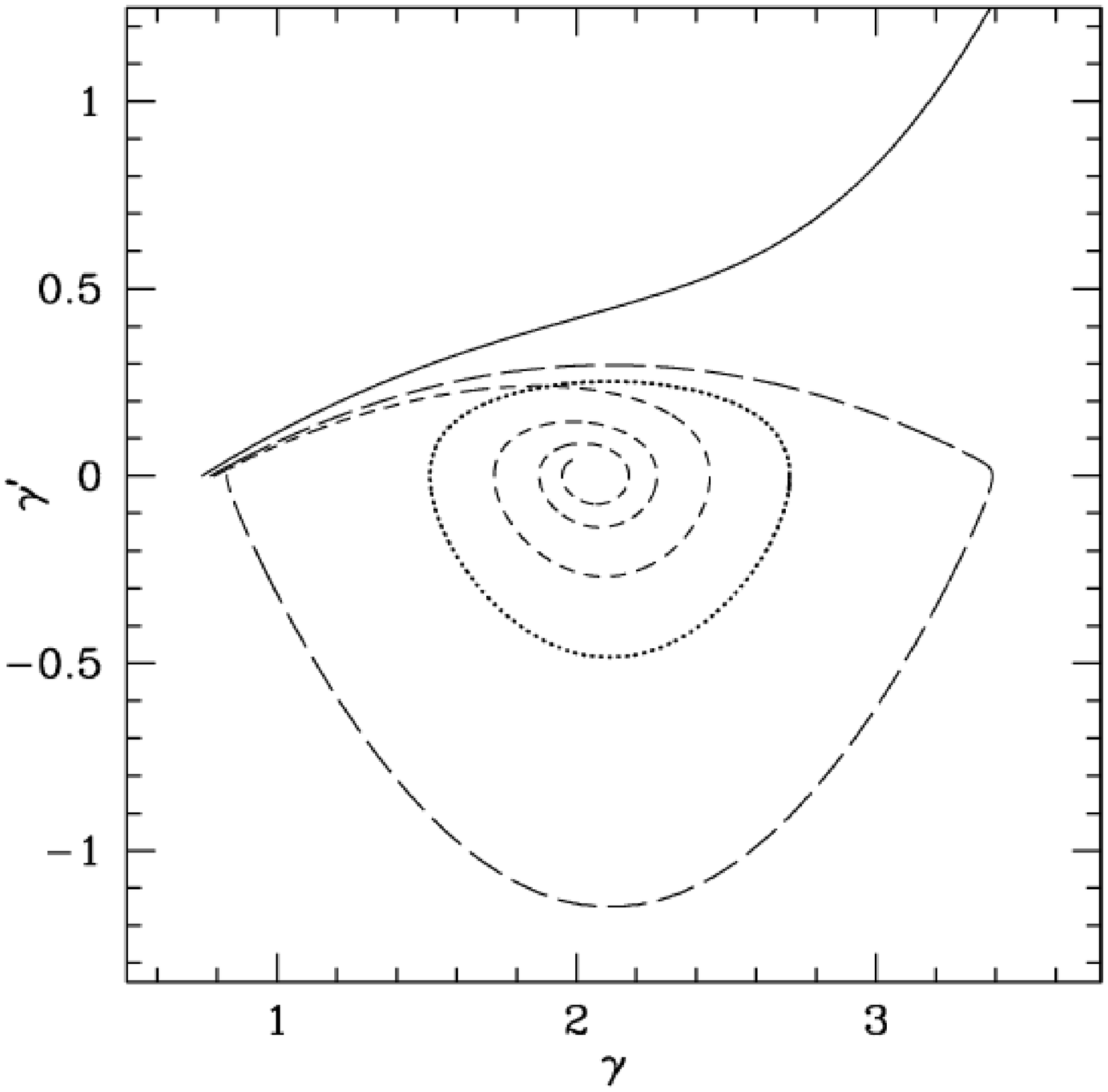} \caption{$\gamma\p$ vs. $\gamma$ for a few Jeans 
equation solutions. ($\gamma$ is the log-log halo density profile slope.) 
The solid line approximates the NFW fit. The long-dash 
line is the limit cycle for equation~\ref{b1}: $\alpha=1.9444$, initial 
$\gamma_{in}=(2/5)\alpha+0.001=0.7788$, initial $\gamma\p_{in}=0$. The dotted line 
(closed loop) also has $\alpha=1.9444$, but different initial $\gamma_{in}=2.7111$, 
and $\gamma\p_{in}=0$. The solid  ($\alpha=1.875$) and short-dash ($\alpha=1.975$) 
lines are the same as in Figure~\ref{smprofiles}. For both, $\gamma_{in}=(2/5)\alpha$
and $\gamma\p_{in}\,\simgt\,0$. The top arc (parabola) of the long-dash line has an 
analytic description, given by equations~\ref{mf} and \ref{tanh}; the bottom arc 
(parabola)is described by equations~\ref{convexmf} and \ref{convextanh}.}
\label{smlimitcycle}
\end{figure}

Thus, $\alpha=1.9444$ results in periodic solutions, and Figure~\ref{smtriangle} 
tells us that this $\alpha$ value is unique.
The amplitude of the wave, i.e. the minimum and maximum values of 
$\gamma$, and its period of oscillations, i.e. the radial span in the halo 
between steepest and shallowest slopes, depend on the initial conditions.
The maximum amplitude is attained for initial $\gamma\p=0$, and initial
$\gamma$ somewhat larger than $(2/5)\alpha$ (lower thick solid line in
Figure~\ref{smtriangle}), or somewhat smaller than $\alpha+1.5$ (upper 
thick solid line in Figure~\ref{smtriangle}). 
As the initial $\gamma$ is set closer to 
$\gamma_{pl}=6-2\alpha=19/9\approx2.111$ the amplitude of oscillations 
decreases. For the exact $\gamma_{pl}$ we get
a power-law density profile, as already noted in Section~\ref{betapl}.
The periodic behavior is illustrated in Figure~\ref{smlimitcycle}                       
which shows the limit cycle of equation~\ref{b1}, as long-dash and dotted line 
closed curves. For both of these $\alpha=1.9444$ and initial $\gamma\p=0$, 
but initial $\gamma_{in}$'s are different (0.7788 and 2.7111). 

\begin{figure}[t]
\epsscale{0.7}
\plotone{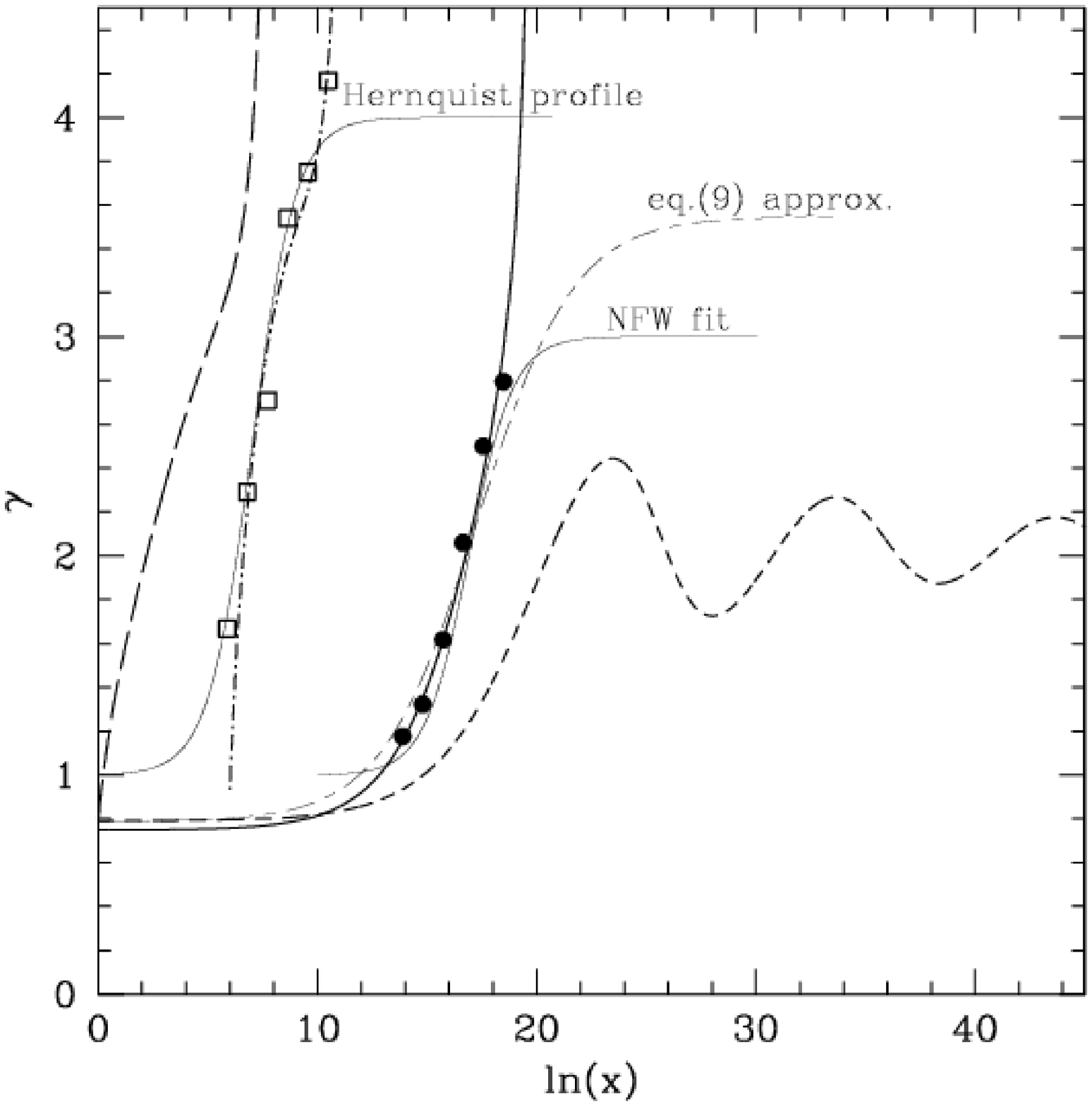}
\caption{Double logarithmic density profile slope, $\gamma$ vs. radius, $x$
(arbitrary normalization). The thick solid line is the solution of the Jeans 
equation that best approximates the NFW fit: $\alpha=1.875$, $\gamma\p_{in}\,\simgt\,0$. 
Short-dash ($\alpha=1.975$, $\gamma\p_{in}\,\simgt\,0$), 
long-dash  ($\alpha=1.975$, $\gamma\p_{in}=1.05$) and 
dot-dash   ($\alpha=2.27$,  $\gamma\p_{in}=2.56$) lines are other Jeans equation 
solutions. All four solutions have $\gamma_{in}=(2/5)\alpha$; 
see Sections~\ref{periodic} and \ref{twoclasses} for details. The solid and short-dash 
lines are the same as in  Figure~\ref{smlimitcycle}. The empty squares and solid 
dots represent Models D (Hernquist) and F (NFW) of \cite{kazan04}. The two thin 
solid lines are the Hernquist (left) and NFW (right) density profile fits. 
}
\label{smprofiles}
\end{figure}

The shape of the upper portion of the $\alpha=1.9444$, 
$\gamma_{in}\approx(2/5)\alpha$ (=0.7788) curve in Figure~\ref{smlimitcycle} 
(long-dash line) suggests that the corresponding solution of the Jeans equation 
has a simple analytical form---a parabola.
In fact, equation~\ref{b1} does admit the following solution:
\begin{equation}
\gamma\p=\gamma\p_m-{1\over 6}(\gamma-\gamma_0)^2,
\quad\quad
\gamma_0={1\over{15}}(55-12\alpha),
\quad\quad
\gamma\p_m={\gamma_0^2\over{6}}+{{2\alpha(\alpha-3)(2\alpha+3)}\over{5(6\alpha+1)}}.
\label{mf}
\end{equation}
Equations~\ref{mf} allow us to express the density 
profile slope as a function of radius in the halo:
\begin{equation}
\gamma=(6\gamma\p_m)^{1/2}\;\tanh\;[(\gamma\p_m/6)^{1/2}\ln x]+\gamma_0.
\label{tanh}
\end{equation}
This solution is valid exactly only for the halo with $\alpha=1.9444$ and 
asymptotic density slopes, $\gamma_{in}=0.7778$ and $\gamma_{out}=3.444$, given 
by the intersection of the $\alpha=1.9444$ line with the upper and lower thick 
solid lines in Figure~\ref{smtriangle}, respectively. For $\alpha$'s within 
$\sim1\%$ of 1.9444 equations~\ref{mf}-\ref{tanh} are a good approximation to
the actual solutions of the Jeans equation. For other values, the approximation
is less valid. For example, for $\alpha=1.875$ the above expression is shown 
as the thin long-short-dash line in Figure~\ref{smprofiles}.                      
For reference, we plot the value of $\gamma_0$, the slope where the density 
slope changes most rapidly, as a thin solid line in Figure~\ref{smtriangle}. 
As expected, for $\alpha=1.9444$, $\gamma_0=\gamma_{pl}=2.111$. 

Finally, we note that $\alpha=1.9444$ periodic solution value 
applies to the case of isotropic velocity dispersion, $\beta=0$. If a halo is 
characterized by a larger, but constant anisotropy, the periodic $\alpha$
value decreases. In the limiting case of radial orbits and $\beta=1$,
$\alpha=1.7222$. Dark matter halos with non-zero anisotropy will be
discussed in a future paper.

\subsection{Two classes of solutions}\label{twoclasses}

It is clear from Figure~\ref{smlimitcycle} that $\alpha=1.9444$ is the 
watershed value that separates two different types of Jeans equation 
solutions: an example of $\alpha>1.9444$ is the short-dash line spiral
curve that eventually settles to a constant finite value of 
$\gamma$; an example of $\alpha<1.9444$ is the solid line which shows a 
rapid increase of $\gamma$ to very large values. The damping term line, 
$8\alpha-5\gamma-5=0$ in Figure~\ref{smtriangle} explains this behavior. 
Solutions with $\alpha<1.9444$ tend to have a negative average damping term
(and density profiles that steepen without limit), while solutions with 
$\alpha>1.9444$ tend to experience positive average damping (and result in
density profiles that asymptote to power-laws, albeit at very large radii). 
We will call these the negative and positive damping solutions, respectively.

Let us consider negative damping solutions and, for concreteness, 
set $\alpha=1.875$. We integrate equation~\ref{b1} from the center of a 
halo outward, i.e. from small to large values of $\ln x$. We start 
from $\gamma_{in}$ given by $2\alpha-5\gamma=0$, and $\gamma\p$ slightly 
greater than 0, which ensures that the center of the halo will have 
constant density profile slope, $\gamma_{in}=(2/5)\alpha$, and hence
vanishing pressure gradient (Section~\ref{alphabeta}).  As $\ln x$ 
increases, $\gamma$ increases without limit, because 
of the negative average damping term.  The solid thick line in 
Figure~\ref{smprofiles} shows the resulting profile. The same profile 
is also shown in Figure~\ref{smlimitcycle}
with the same line type. (Note that the horizontal normalization is not 
important, and was adjusted for easier comparison between various results.)

To illustrate the behavior of positive damping term solutions we set
$\alpha=1.975$.  As before, we choose $\gamma_{in}=(2/5)\alpha$, and 
initial $\gamma\p$ slightly greater than 0.
This solution is plotted as the short-dash line in Figures~\ref{smlimitcycle} 
and \ref{smprofiles}.
It does not attain steep density slopes in the outer halo; $\gamma$ climbs
to about 2.45, then oscillates, with decreasing amplitude, and eventually 
settles to the asymptotic value given by $2\alpha+\gamma_{pl}-6=0$.
It is the positive average damping term in equation~\ref{b1} that forces 
the oscillations to a constant asymptotic $\gamma_{pl}$.
The halo represented as the short-dash line has an outer density slope
$\gamma<3$, implying that its mass does not converge. However, this is 
not a problem because in a real halo the profile pictured here would not 
apply at arbitrarily large radii where the assumptions of the Jeans 
equation will break down, including the condition of equilibrium itself. 
A real halo will have a steep outer slope, and convergent total mass.

Thus, the behavior of $\gamma$ vs. $\ln x$ for $\alpha$ greater and less
than the periodic value (1.9444) are qualitatively different. If $\alpha<1.9444$ 
the outer density slope increases without limit, while if $\alpha>1.9444$ 
there is a maximum value for $\gamma$, followed by oscillations. 

If we abandon the $\gamma\p_{in}\;\simgt\;0$ initial condition and instead use
$\gamma\p_{in}>0$ we can ``overpower'' the positive damping term and make
$\gamma$ attain large values. The solution obtained with $\gamma\p_{in}=1.05$ 
and $\alpha=1.975$ is plotted as the long-dash line in Figure~\ref{smprofiles}.  
The price we pay for resorting to large $\gamma\p_{in}$ is non-zero
${d\ln(\rho\sigma^2)}/{d\ln r}$  at the center of the halo.
(Furthermore, $\gamma$ decreases towards the center, and becomes negative:
these halos have density holes in the center.) It is impossible to make a 
halo with $\alpha>1.9444$, constant pressure core, and outer halo $\gamma>3-4$ 
within the confines of the spherically symmetric, isotropic velocity 
dispersion Jeans equation.

\subsubsection{NFW vs. Hernquist-type density profiles}
\label{HernvsNFW}

We illustrate the positive and negative damping solutions with Hernquist 
and NFW density profiles, respectively. The recent work of \cite{kazan04} 
provides us with N-body realizations of Hernquist (their Model D) and NFW 
(Model F) profiles. These halos are not the result of the usual cosmological
evolution starting from small density perturbations; instead they were
explicitly constructed as isolated equilibrium objects.
To qualify for the analysis using equation~\ref{b1} the halos 
must be in equilibrium, have isotropic velocity dispersion ellipsoids, 
and power-law phase-space density distributions. Models D and F meet the
first two requirements almost exactly, while the last requirement is
met approximately, as we describe below.
The N-body generated halos have been evolved for 50 
crossing times and thus have been explicitly shown to be in equilibrium. 
The velocity ellipsoid parameter, $\beta$, equation~\ref{betaell}, 
is zero for Models D and F. To estimate the phase-space density distribution 
we used Figure~3 of \cite{kazan04} to measure density and velocity dispersion 
profiles at seven points over 2-2.5 decades in radius. 

The phase-space density distribution of Model D is reasonably approximated 
by a power-law, but it is not an exact power-law: 
the slope based on the inner half of the halo is $\alpha=2.01$, 
while the outer half has $\alpha=2.53$. The average over the whole halo is 
with $\alpha=2.27\pm0.11$; the error is the RMS dispersion 
about the best fit line. We use this $\alpha$ value to generate the Jeans 
equation solution, and plot it as the dot-dash line in Figure~\ref{smprofiles}. 
The solution and the actual data points from \cite{kazan04} (empty squares)
are in very good agreement. For reference, we also plot the Hernquist 
density profile, $\rho\propto [{x(1+x)^3}]^{-1}$, as a thin solid line.

The phase-space density profile of Model F is well fit by a single 
power-law with $\alpha=1.95\pm0.04$. This value is in excellent agreement 
with the periodic solution $\alpha=1.9444$ that we derived from the Jeans 
equation in Section~\ref{periodic}. Since \cite{kazan04} did not set out
to produce halos with $\alpha$ close to the periodic value, we speculate
that 1.9444 is an attractor, albeit a weak one:  halos whose properties are 
already close to those of the periodic solution (because of input constraints, 
dynamical evolution, or some other reason), end up with $\alpha\approx1.9444$. 
The range of $\alpha$ values exhibited by the ESIM halos in Section~\ref{ESIMsec}
also attests to the weakness of the ``periodic solution'' attractor.

The density profile of Model F (solid dots in Figure~\ref{smprofiles}) can 
be reproduced with the Jeans equation solution with $\alpha=1.95$ and large
$\gamma\p_{in}$, around 1, or with a smaller $\alpha$ and $\gamma\p_{in}=0$. 
For example, $\alpha=1.875$ and $\gamma\p_{in}=0$ solution is plotted as
a thick solid line; the NFW empirical fit (thin solid line) is plotted
for reference.

While the Model F value of $\alpha=1.95\pm0.04$ is fully consistent with the 
periodic solution, it is inconsistent with TN01's 1.875 at $\sim 2\sigma$. 
There are at least two possibilities why the $\alpha$ value obtained from 
N-body simulations (1.875) deviates from the periodic solution (1.9444) 
obtained assuming symmetric, isotropic halos in perfect hydrostatic equilibrium. 
N-body halos are subject to dynamical evolution, like gravitational collapse 
and mergers. Any of these could drive $\alpha$ away from the periodic solution 
value. (We remind the reader that Kazantzidis et al. models did not undergo 
the type of dynamical evolution that cosmological halos undergo.)
Alternatively, it is possible that our assumption of zero velocity dispersion
anisotropy is not a good enough approximation. The actual anisotropy in halos 
has a gradient, with the inner region being isotropic and outer regions being 
more radial \citep{barnes05a,hm04,cl96}. Extending equation~\ref{b1} from 
$\beta=0$ to constant positive anisotropy $\beta\ne 0$, shifts the periodic 
solution to smaller $\alpha$ values. Specifically, $\alpha=1.875$ is the 
periodic solution for $\beta=0.3125$ equilibrium halo.  
We postpone a detailed treatment of velocity anisotropy to a later paper.
Finally, there are other, more subtle issues that can drive $\alpha$ away
from its periodic value; halo triaxiality is one example. 

In this subsection we discussed \cite{kazan04} models as examples of 
equilibrium halos and compared them to the results of the Jeans equation 
analysis. However, these halos may not be of much help when trying to 
understand halos formed in cosmological simulations, because of the very 
different dynamical evolution that the latter undergo. In other words, 
not all halos that are in equilibrium could have been formed in 
cosmological settings. A good example is provided by the Hernquist 
profile (Model D), which is in equilibrium, but does not describe halos 
generated in cosmological simulations. It follows that the values of 
$\alpha$ measured from the Hernquist profile are not particularly 
useful in elucidating the nature of halos evolved in cosmological
settings, either N-body or ESIM.

\subsection{Conclusions from the Jeans equation analysis}

Our goals in Section~\ref{jeans} were two-fold: gain insight into the 
properties of solutions of the Jeans equation, and find out if the
Jeans equation singles out any value of $\alpha$.
Analyzing equation~\ref{b1} we found that the isotropic Jeans equation 
does indeed point to a special $\alpha$: $\alpha=1.9444$ results in 
periodic variation of $\gamma$ vs. $\ln x$. Thus, $\alpha=1.9444$ divides 
the solutions into those where the damping term is typically negative, and 
$\gamma$ increases without limit, producing halos with density profiles 
that steepen at large radii, and those where the damping term is typically 
positive, and $\gamma$ asymptotes to a constant value, and hence power-law 
density profiles (provided the initial $\gamma\p$ is small). Interestingly,
for the unique periodic solution $\alpha=1.9444$ the shallowest inner
density slope and the steepest outer slope are 0.78 and 3.44 respectively,
values which are similar to those of N-body generated halos.

For halos with constant non-zero velocity anisotropy, $\beta$, the periodic
solution shifts to smaller values of $\alpha$, such that halos with purely 
radial orbits have periodic $\alpha=1.7222$. The N-body simulated
halos have average anisotropies between 0 and 1; so the corresponding
periodic solution $\alpha$ should be bracketed by 1.7222 and 1.9444. Given 
this information, it is not surprising that the N-body generated halos
that have undergone considerable dynamical evolution and have $\beta>0$ 
are characterized by an intermediate $\alpha$ value, $\alpha\approx1.88$.
Our own tests with ESIM halos (Section~\ref{ESIMsec}) produce a range of 
$\alpha$ values, depending on the parameters of the initial set-up, and
the properties of final halos. In a sense, ESIM halo results agree with 
the conclusions from the Jeans equation analysis because neither predict 
that all halos should have the same $\alpha$. Based on the analysis of 
Section~\ref{ESIMsec} and \ref{jeans} we conclude that the radial slope 
of the phase-space density, $\alpha$, is not fixed by the condition of 
hydrostatic equilibrium, and so must be the result of some other physics.

\section{Conclusions}

Examination of a handful of N-body dark matter halos indicates that the 
phase-space density, \powerlaw, is a power-law in radius, even though the 
density profile is not \citep{tn01}.  To determine if the scale-free 
nature of \powerlaw\/ is peculiar to N-body halos (may be even just those
presented in TN01), or is a more generic outcome of collisionless 
gravitational collapse, we turned to a different method of generating 
equilibrium halos. The Extended Secondary Infall Model 
(ESIM) halo formation, which we used in Section~\ref{ESIMsec}, proceeds
very differently from that of N-body simulations. ESIM halos are 
spherically symmetric, and non-rotating. By assumption, the collapse 
conserves the angular and radial momenta of all shells individually.
The final equilibrium halos are supported against collapse by radial and
tangential velocity dispersion.

Despite the differences between N-body and ESIM methods, and the resulting 
density profiles, both types of halos are characterized by a scale-free 
\powerlaw\/ profile. We speculate that because this feature is common to the two 
types of halos, it is an outcome of the one piece of physics shared by 
the two methods: violent relaxation. 
At the very least violent relaxation is a sufficient, if not the 
necessary condition needed to produce \powerlaw$\propto r^{-\alpha}$.
We will further test this hypothesis in subsequent work.

If the scale-free nature of \powerlaw\/ is, in fact, a generic feature of 
collisionless halos, it implies an additional constraint on the halo structure 
as the density and velocity dispersion profiles are now closely related.
In particular, the analysis of the Jeans equation and subsequent 
implications for halo formation are now not as unconstrained as before. 
Using this additional condition, we have rewritten the second derivative of 
the Jeans equation as equation~\ref{b1}, and used it as the starting point 
of our analysis in Section~\ref{jeans}. This analysis yielded some interesting 
and novel insights into the solutions of the Jeans equation. In particular, 
we identified the unique periodic solution of equation~\ref{b1}, where the
halo density profile slope oscillates between a minimum values of 0.78 and 
a maximum value of 3.44. These density profiles slopes are close to the
values found in N-body simulated halos. 
The corresponding $\alpha=1.9444$ is the only special $\alpha$ value admitted
by the spherically-symmetric, isotropic Jeans equation.

In the future, we plan to extend the analysis of the Jeans equation to include 
non-zero velocity dispersion anisotropy distributions. Our ultimate goal is to 
identify the physical processes that shape the dark matter
halos produced in simulations, and eventually, in the real Universe.

\acknowledgements
This work was supported in part by NSF grant AST 03-07604.
AB also receives research support from the Natural Sciences and 
Engineering Research Council (Canada) through the Discovery grant program. 
He further acknowledges support from the Leverhulme Trust (UK) in the form 
of the Leverhulme Visiting Professorship at the Universities of Oxford 
and Durham, where part of this work was carried out. 
JJD was partially supported through the Alfred P.\ Sloan Foundation.
The authors are grateful to Adi Nusser for his insightful comments.
LLRW would like to thank Philip C. Williams for enlightening 
discussions about non-linear oscillators.

\appendix
\section{Convex density profile solutions of the Jeans equation}\label{convex}

All analytical density profiles that appear in the literature and describe mass 
(or light) distribution in galaxies are ``concave'', i.e. the double logarithmic 
density slope steepens with radius, $\gamma\p>0$. However, the Jeans equation 
also admits solutions with $\gamma\p<0$. These can be obtained by integrating 
equation~\ref{b1} starting with $\gamma_{in}$ given by $\tth[\alpha-\gamma]+1=0$ and 
$\gamma\p_{in}\;\simlt\; 0$. The density profiles asymptote to constant 
$\gamma_{in}=\alpha+1.5$ at the center, and constant $\gamma_{out}=(2/5)\alpha$ 
in the outer parts. In the representation of Figure~\ref{smtriangle} these 
profiles start at the upper thick solid line and end at the lower thick solid 
line. For any value of $\alpha$ the exact analytical solution is,
\begin{equation}
\gamma\p=-\gamma\p_m+\tth(\gamma-\gamma_0)^2,
\quad\quad
\gamma_0=(14\alpha+15)/20,
\quad\quad
\gamma\p_m=3(2\alpha+5)^2/200,
\label{convexmf}
\end{equation}
and the double logarithmic density profile slope is given by
\begin{equation}
\gamma=-(3\gamma\p_m/2)^{1/2}\;\tanh\;[(2\gamma\p_m/3)^{1/2}\ln x]+\gamma_0.
\label{convextanh}
\end{equation}
The form of these equations is similar to that of equations~\ref{mf}-\ref{tanh} 
that describe concave density profiles (Section~\ref{periodic}). 
The value of the density profile slope where the density slope changes most 
rapidly, $\gamma_0$, is depicted by a thin solid line in Figure~\ref{smtriangle}.
An example of this type of
density profile is the lower half ($\gamma\p<0$) of the long dash line in 
Figure~\ref{smlimitcycle}. In these halos the velocity dispersion, $\sigma$, 
after reaching its minimum value at the radius where $\gamma=\gamma_0$, starts 
to increase in the outer parts of the halo. The pressure, $\rho\sigma^2$, 
asymptotes to a constant in the outer parts
because $\gamma_{out}=(2/5)\alpha$ is constant there
(see Section~\ref{alphabeta}).

\section{Phase-space density of halos and the secondary infall model}
\label{SIM}

In Section~\ref{ESIMsec} we examined the phase-space density of halos
formed via the Extended Secondary Infall Model, ESIM, which differs from 
the original, Secondary Infall Model, SIM \citep{gg72,g75} in many respects. 
The most notable differences are the shape of the pre-collapse peak
(extended in ESIM, point mass in SIM), and the inclusion of secondary 
perturbations, and hence random motions in the ESIM (SIM halo collapse is 
purely radial). With these differences in mind, it is interesting to compare
$\alpha$ values obtained in Section~\ref{ESIMsec} to the prediction of SIM.

In a spherically symmetric SIM halo collapse proceeds as follows.
Initially, a small constant central mass excess, $\delta M_0$ is surrounded 
by material of average density. The dynamics of the pre-turn-around period 
is described by the parametric equations, $r\propto (1-\cos\theta)$, and 
$t\propto (\theta-\sin\theta)$. The maximum radius for a shell of initial 
comoving radius $r_i$ is $r_m=r_i\bar\delta^{-1}$, where $\bar\delta$ is 
the initial fractional overdensity inside $r_i$. Enclosed mass, 
$M(<r_i)\propto\rho r_i^3$, so $r_i\propto M^{1/3}$. And, 
$\bar\delta=\delta M_0/M\propto M^{-1}$.  
Combining these scalings we get $r_m\propto M^{4/3}$, for each shell in
the halo. After reaching turn-around, a shell collapses back somewhat; 
it is usually assumed that the collapse factor, i.e. the ratio of the 
turn-around radius, $r_m$, to the final radius in the virialized halo, 
is the same for all shells. Further assuming that each shell spends most 
of the time at its apocenter, which is proportional to $r_m$, one 
derives the resulting density distribution in the final virialized halo:
$\rho(r)\propto r^{-2}\,dM/dr$, or, ~$\rho\propto M^{-3}\propto r_m^{-9/4}$. 
This is a well known result.  

Let us take this approach a little further. In the real Universe the 
collapse of material will not be purely radial; there will be some random 
motion of particles, and associated kinetic energy. If we assume that 
during the collapse the kinetic energy is derived from the gravitational 
potential energy of infalling material, then the two types of shell energy
are proportional, and ${1\over 2} v^2\propto GM/r_m$.  Kinetic energy of 
random motion gives an estimate of the velocity dispersion: 
$\sigma\propto (M/r_m)^{1/2}\propto r_m^{-1/8}$. Combining $\rho(r)$ from the 
previous paragraph with $\sigma(r)$, we get the radial dependence of the 
phase-space density, $\rho/\sigma^3\propto (r^{-9/4})/(r^{-3/8})\propto r^{-15/8}$,
~i.e. $\alpha=1.875$. If the kinetic energy is not simply proportional
to the potential energy, the final value of $\alpha$ will differ accordingly.
The SIM phase-space density shape is always a power-law, by definition. 

These are the predictions of the highly simplified SIM. Given that ESIM halos
of Section~\ref{ESIMsec} have more complexity, and a variety of initial 
conditions, it is not surprising that ESIM halos show a wide range of $\alpha$
values. However, the typical values are not far from the SIM prediction of
1.875.  Finally, the fact that $\alpha=1.875$ derived here, has
the same numerical value as the best-fit $\alpha$ quoted by \cite{tn01} for
their N-body generated halos, is probably a coincidence, because the 
hierarchical halo formation characteristic of N-body simulations is 
qualitatively different from the highly idealized SIM.

\end{document}